\documentclass[aps,prl,twocolumn,superscriptaddress,showpacs]{revtex4-1}

\usepackage{amsmath,amssymb,graphics,epsfig,epstopdf,color,verbatim,ulem,braket,tabularx,graphicx,epsfig,subfigure,psfrag}
\usepackage[colorlinks,linkcolor=blue,citecolor=blue,urlcolor=blue]{hyperref}

\begin{document}
\title{Dirac and nodal line magnons in three-dimensional antiferromagnets}
\author{Kangkang Li}
\thanks{These authors contributed equally to this study.}
\affiliation{Beijing National Laboratory for Condensed Matter Physics,
and Institute of Physics, Chinese Academy of Sciences, Beijing 100190, China}
\affiliation{University of Chinese Academy of Sciences, Beijing 100049}
\author{Chenyuan Li}
\thanks{These authors contributed equally to this study.}
\affiliation{International Center for Quantum Materials, School of Physics, Peking University, Beijing 100871, China}
\author{Jiangping Hu}
\affiliation{Beijing National Laboratory for Condensed Matter Physics,
and Institute of Physics, Chinese Academy of Sciences, Beijing 100190, China}
\affiliation{Collaborative Innovation Center of Quantum Matter, Beijing 100871, China}
\author{Yuan Li}
\email{yuan.li@pku.edu.cn}
\affiliation{International Center for Quantum Materials, School of Physics, Peking University, Beijing 100871, China}
\affiliation{Collaborative Innovation Center of Quantum Matter, Beijing 100871, China}
\author{Chen Fang}
\email{cfang@iphy.ac.cn}
\affiliation{Beijing National Laboratory for Condensed Matter Physics,
and Institute of Physics, Chinese Academy of Sciences, Beijing 100190, China}
\begin{abstract}
We study the topological properties of magnon excitations in three-dimensional antiferromagnets, where the ground state configuration is invariant under time-reversal followed by space-inversion ($PT$-symmetry). We prove that Dirac points and nodal lines, the former being the limiting case of the latter, are the generic forms of symmetry-protected band crossings between magnon branches. As a concrete example, we study a Heisenberg spin model for a ``spin-web'' compound, Cu$_3$TeO$_6$, and show the presence of the magnon Dirac points assuming a collinear magnetic structure. Upon turning on symmetry-allowed Dzyaloshinsky-Moriya interactions, which introduce a small non-collinearity in the ground state configuration, we find that the Dirac points expand into nodal lines with nontrivial $Z_2$-topological charge, a new type of nodal lines unpredicted in any materials so far.
\end{abstract}
\maketitle
\textit{Introduction} The theoretical proposal \cite{Murakami2007,Wan2011} and experimental discovery \cite{Lv2015,Xu2015a,Lu2015} of Weyl semimetals have opened up a new research field called topological semimetals\cite{Burkov2016}.
Physically, the essence of topological band theory, that the Bloch wavefunction on a closed surface in momentum space can have nontrivial topological structures, is independent of the statistics of the constituent particles \cite{Haldane2008,Lu2014,Huber2016}.
By replacing the electronic spin polarization in the above example by light polarization, for instance, one obtains a topological band crossing in photonic crystals.
Such ideas of generalization have inspired researchers to find topologically nontrivial band crossings in boson systems of photons \cite{Lu2013,Wang2016}, phonons\cite{Stenull2016} and magnons in three\cite{Li2016} and lower dimensions\cite{Fransson2016,Owerre2016a,Owerre2016}.

Topological classification solely depends on symmetry class and dimensionality.
There have been many studies on topological band crossings protected by lattice space group symmetries\cite{Young2012,Wang2012,Wang2013,Yang2014,Fang2016,Bradlyn2016,Weng2016,Zhu2016,Chang2016,Wieder2016,Yang2017,Wieder2017}, and a (almost) full classification of this type has appeared in the literature\cite{Watanabe2016}.
In this paper, we focus on a new type of symmetry groups: the magnetic groups, which naturally rise in magnetically ordered systems.
The difference between a magnetic group and a space group is that the former generically contains elements of the form $ST$, where $S$ is some space-group operation and $T$ time-reversal \cite{Bradley1968}, while neither $S$ nor $T$ is a symmetry.
Band crossings protected by magnetic groups can be found in the electronic band structures in magnetic materials, as well as the band structure of magnons (coherent spin excitations) over a magnetic ground state.

We choose for our study one of the simplest magnetic groups, generated by $PT$, where $P$ is spatial inversion.
This magnetic group pertains to various antiferromagnets with centro-symmetric crystal lattices, where two spins related by inversion have opposite polarizations in the ordered state.
We first study the case where there is, in addition to $PT$, a global U(1) spin-rotation symmetry, seen in most collinear antiferromagnets with or without an easy axis.
We find that in the spin wave dispersion the generic band crossings among the magnon branches are Dirac points having integer monopole charges.
Furthermore, we find that when the U(1) symmetry is broken (\textit{e.g.}, by Dzyaloshinsky-Moriya interactions (DMI)\cite{Dzyaloshinsky1958,Moriya1960} or other anisotropic effects), but $PT$ still preserved, each Dirac point \textit{necessarily} becomes a nodal line.
Unlike all nodal lines so far predicted in materials\cite{Fang2016a}, this nodal line cannot continuously shrink to a point and disappear, because it is protected by a new $Z_2$ monopole charge\cite{Fang2015a}, aside from the $\pi$-Berry phase common to all nodal lines\cite{Burkov2011,Xu2011,Volovik2013,Kim2015,Fang2015a}.
We apply the general theory to a three-dimensional ``spin-web'' compound, Cu$_3$TeO$_6$\cite{Herak2005,Choi2008}, which develops a long-range and almost collinear antiferromagnetic order below $T_\mathrm{N}\approx61$ K.
We use a $J_1$-$J_2$ ($J_1>J_2>0$) Heisenberg model to describe the spin interactions and calculate the magnon band structure using linear-spin-wave approximation, where multiple pairs of Dirac points are identified between two optical magnon branches.
Then we add Dzyaloshinsky-Moriya interactions to the model, and calculate the new classical ground state as well as the spin wave excitations over the non-collinear ground state.
Comparing with the previous results, we find that each Dirac point now becomes a nodal line, whose size is proportional to the strength of DMI squared.
Experiments for detecting key features of Dirac and nodal line magnons are proposed.

\textit{General theory} We begin by noting that when the total $S_z$ is preserved (\textit{i.e.}, with U(1) symmetry), all single-particle excitations can be labeled by their spin quantum numbers.
For magnons, these numbers are $+1$ and $-1$,  and magnons with opposite spins are decoupled in a quadratic Hamiltonian.
Next we look at how the magnetic-group symmetry $PT$ acts on the magnons.
Physically, spatial-inversion preserves spin and time-reversal inverts it, making the composite symmetry $PT$ invert the spin quantum number of a magnon.
Based on these observations, we see that the single-particle Hamiltonian decouples into two sectors, one for each spin quantum number, or symbolically
\begin{equation}
H=H_+\oplus{H}_-,
\end{equation}
where $H_\pm$ is the Hamiltonian for the spin-$\pm1$ sector in the spin system.
Magnetic group symmetry $PT$ then requires
\begin{equation}\label{eq:3}
H_+=H^\ast_-.
\end{equation}
When $H_+$ is defined in three-dimensional Brillouin zone, there can be Weyl points in the spectrum as band crossings\cite{Murakami2007}.
Due to Eq.~(\ref{eq:3}), $H_-$ and $H_+$ have the same band structure and therefore when a Weyl point appear in $H_+$, there must be another Weyl point in $H_-$ at the same (crystal) momentum.
Since $PT$-symmetry reverses the Berry curvature, these two Weyl points are of opposite monopole charge, so together they make a Dirac point.
Below are some basic properties of these Dirac points in the bulk.
(i) The Weyl points in $H_+$ are not pinned to any high-symmetry point, line or plane, so the Dirac point may appear at any momentum in the Brillouin zone, in contrast to previously studied Dirac points that are pinned to high-symmetry points and lines.
(ii) Since the Weyl points in $H_+$ must appear in pairs, so do the Dirac points in $H$.
(iii) For each Dirac point we can define a monopole charge as the monopole charge of the associated Weyl point in $H_+$, a $Z$-index.

In realistic magnetic materials, besides the isotropic Heisenberg terms, other terms, such as site-dependent single-ion anisotropy and exchange anisotropy, may break the U(1) spin-rotation symmetry but will leave the space-time symmetry $PT$ intact (or there would be ferroelectricity).
For example, when the bonds connecting two magnetic atoms to their common ligand atom make an angle less than 180$^\circ$, DMI is in general present.
When U(1) symmetry is broken, a Dirac point is no longer stable, and as long as $PT$ is still preserved, Weyl points are disallowed \cite{Murakami2007}, so in principle a Dirac point must be either fully gapped out or broken into a nodal line.
Further analysis rules out the former possibility, and shows that each Dirac point becomes a nodal line upon turning on these anisotropic perturbations.
To see this, we notice that as long as $PT$-symmetry is preserved, even in the absence of U(1)-symmetry, a $Z_2$ topological invariant can still be defined on a sphere surrounding the Dirac point, which is found to be nontrivial [see Ref.[\onlinecite{SupMat}] for calculation] for any sphere containing one (or an odd number of)  Dirac point(s). According to Ref.[\onlinecite{Fang2015a}], the nontrivial invariant indicates that the Dirac point is but a limiting case of a nodal line, which cannot be gapped out as long as $PT$ is preserved.
To our best knowledge, while nodal lines without $Z_2$ monopole charge have been proposed in many fermionic and bosonic systems\cite{Fang2016a}, nodal lines carrying nontrivial $Z_2$ monopole charge have so far not been predicted in any real materials.

\begin{figure}
\includegraphics[width=8cm]{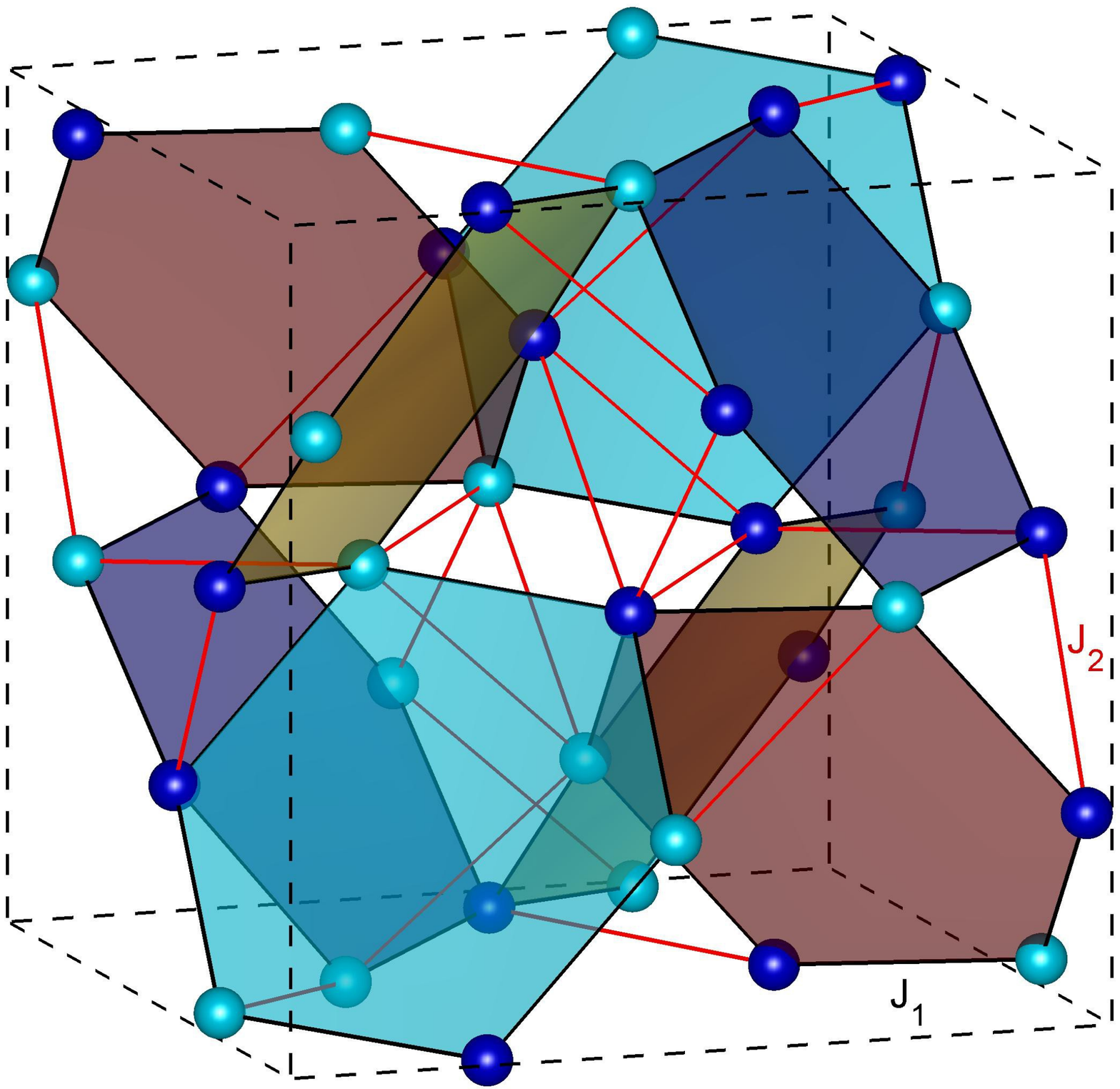}
\caption{The Cu$^{2+}$ sublattice of Cu$_3$TeO$_6$ in a cubic unit cell, with spin-up and -down ions represented in different colors. Nearest-neighbor ($J_1$) and next-nearest-neighbor ($J_2$) interactions are indicated.}
\label{fig:structure}
\end{figure}

\textit{Dirac magnons in Cu$_3$TeO$_6$}
Three-dimensional collinear antiferromagnets are the best platform for us to realize these topological band crossings in $k$-space. Here we have chosen Cu$_3$TeO$_6$, which was reported to host a novel spin lattice\cite{Herak2005}, dubbed a three-dimensional spin web\cite{Choi2008,Mansson2012}.
The lattice consists of almost coplanar Cu$^{2+}$ hexagons that are perpendicular to one of the four space diagonals of the cubic unit cell (Fig.~\ref{fig:structure}), featuring a hybrid between a 3D spin-1/2 network and a low connectivity of interactions between neighbors: each Cu$^{2+}$ ion is shared by two hexagons and has only four nearest neighbors (and four next nearest neighbors).
Below $T_\mathrm{N} \approx 61$ K, the system develops long-range antiferromagnetic order that leaves clear signatures in magnetic susceptibility and neutron diffraction measurements \cite{Herak2005}. Without loss of generality, we believe that the large yet highly symmetric magnetic primitive cell of Cu$_3$TeO$_6$ is favorable for symmetry-protected magnon band crossings.

Furthermore, we note that the lattice structure of Cu$_3$TeO$_6$ is very similar to those of $C$-type sesquioxides $R_2$O$_3$ ($R =$ Y, Sc, In, or rare-earth elements) \cite{Maslen1996}.
The spin lattice of Cu$_3$TeO$_6$ can be realized in the latter, if the Wyckoff 24$d$ and 8$a$ sites can be occupied by magnetic and non-magnetic ions, respectively.
Given the rather broad distribution of the $R^{3+}$ ionic radii, ranging from 81 pm (Sc and In) to 106 pm (La), it might be possible to synthesize solid solutions of them, such as Nd$_3$ScO$_6$, with minimal inter-site disorder \cite{Antic1995}.
Along with the rich magnetic properties of rare-earth elements, this renders our analysis of Cu$_3$TeO$_6$ potentially applicable to a large family of interesting magnetic materials.

Available neutron diffraction data are consistent with a collinear antiferromagnetic spin configuration depicted in Fig.~\ref{fig:structure}, although a slightly non-collinear tilting cannot be ruled out \cite{Herak2005}.
Within this part, we assume the collinear ground state, \textit{i.e.}, U(1) spin rotation symmetry; and in the next we include Dzyaloshinsky-Moriya interactions to account for the effect of non-collinearity.
The collinear ground state is most easily understood by assuming the unfrustrated nearest-neighbor Heisenberg exchange interaction $J\mathbf{S}_i\cdot\mathbf{S}_j$.
Yet, due to the geometric configuration of the atoms, the next-nearest-neighbor exchange may also have an appreciable magnitude, whose sign is likely to be also positive (antiferromagnetic).
We thus model the spin interactions in Cu$_3$TeO$_6$ using the following $J_1$-$J_2$ Heisenberg model
\begin{equation}\label{eq:4}
H=J_1\sum_{\langle{ij}\rangle}\mathbf{S}_i\cdot\mathbf{S}_j+J_2\sum_{\langle\langle{ij}\rangle\rangle}\mathbf{S}_i\cdot\mathbf{S}_j.
\end{equation}
The classical ground state of $H$ depends on the relative magnitude of $J_1$ and $J_2$, and when $J_2<J_c=J_1/3$, the ground state configuration matches the experimental one shown in Fig.~\ref{fig:structure}.
It is easy to check that this spin configuration preserves both $PT$ and $S_z$, and hence it may host Dirac magnons.
Assuming strongly localized moments and negligible quantum fluctuations, we treat the magnon excitations using the linear spin-wave approximation.
\begin{figure}
\includegraphics[width=8.5cm]{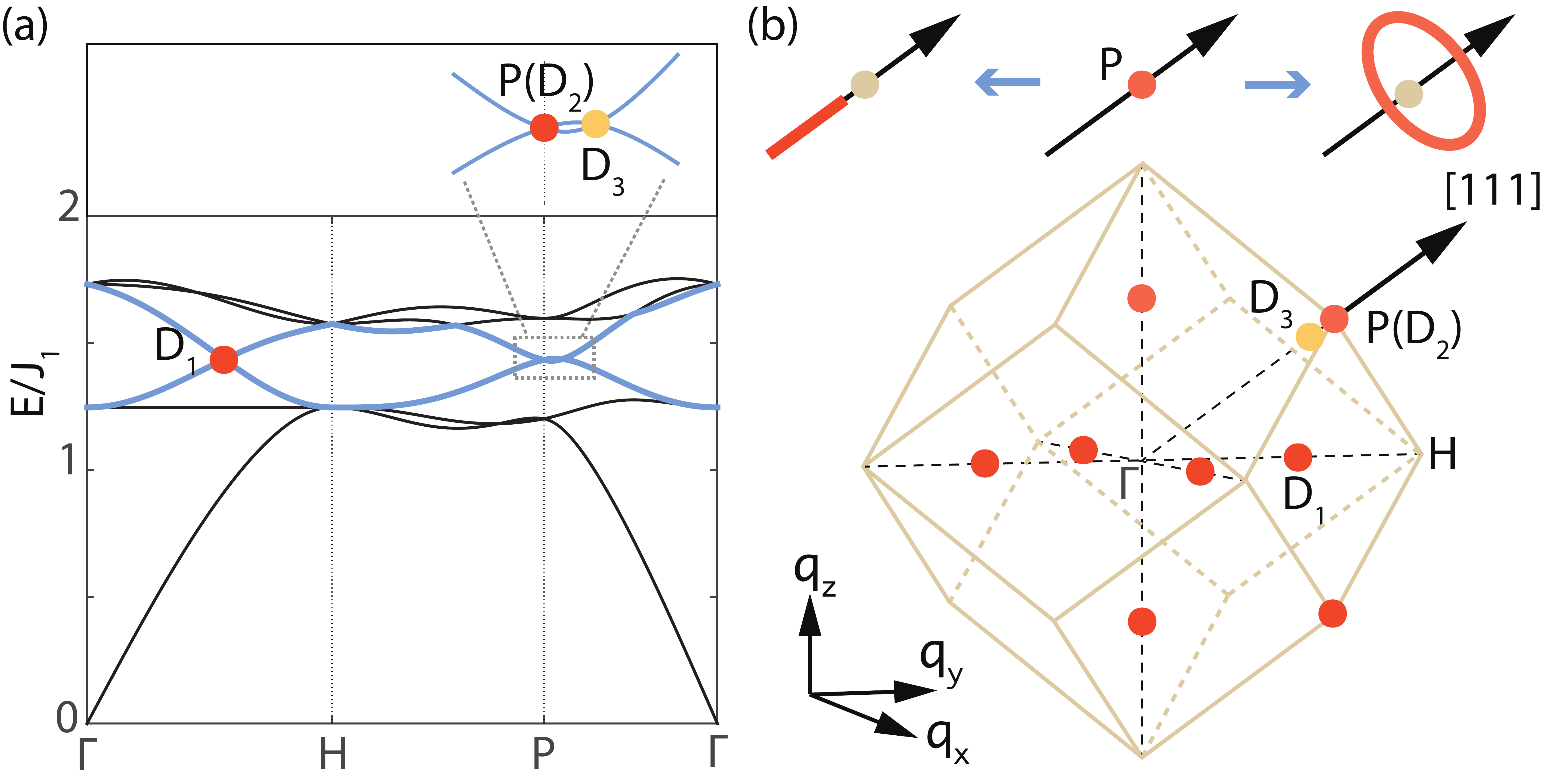}
\caption{(a) A typical band structure of the spin-wave dispersion along high-symmetry lines with $J_2=0.134J_1>0$, where the inset shows a zoomed-in region near P. (b) Positions of all Dirac points in the first Brillouin zone. Red and yellow colors indicate the monopole charge of +1 and -1, respectively. For clarity, only one of the eight D$_3$ points is displayed in the three-dimensional Brillouin zone in (b). Above the Brillouin zone, we schematically show how, upon adding DMI, a Dirac point at P expands either into a nodal ring about [111] or into a line along [111], preserving the threefold rotation along the [111]-axis.}
\label{fig:bands}
\end{figure}
There are twelve spins in each primitive cell, with six pointing positive (along a domain-dependent $\langle111\rangle$-direction \cite{Herak2005}, which we refer to as the [111]-direction) and six negative in the ground state. (Note that the magnetic order does not enlarge the lattice primitive cell.) We perform the standard Holstein-Primakoff transformation on the up spins
\begin{equation}\label{eq:HP1}
S_+=\sqrt{2S}a, S_z=S-a^\dag{a},
\end{equation}
and down spins
\begin{equation}\label{eq:HP2}
S_+=-\sqrt{2S}b^\dag, S_z=-S+b^\dag{b},
\end{equation}
where $S_+\equiv{S}_x+{i}S_y$.
We remark that under spin rotation along $z$-axis through $\theta$, spin wave operator $a$ transforms as $a\rightarrow{a}e^{-i\theta}$ on up spins and $b\rightarrow{b}e^{i\theta}$ on down spins, making them $S_z=+1$ and $-1$ operators, respectively.
All spin-wave operators can thus be divided into two sets by their spins: $\{a, b^\dag\}$ having $S_z=+1$ and $\{a^\dag,b\}$ having $S_z=-1$.
As long as the U(1) symmetry is present, these two sets do not couple to each other in a quadratic spin-wave Hamiltonian. The steps we take to find and solve the spin wave Hamiltonian are given in Ref.[\onlinecite{SupMat}].

For $J_2=0.134J_1$ (but see Ref.[\onlinecite{SupMat}] for other values of $J_2$), the magnon bands along high-symmetry lines in the Brillouin zone are plotted in Fig.~\ref{fig:bands}(a).
Distinct linear band crossings can be found between two optical branches in pale yellow.
Calculation of the monopole charge using the Wilson loop technique confirms that all these band crossings are Dirac points  (or Weyl points in $H_+$): there are six positive Dirac points along $\Gamma$H and its symmetry equivalents (denoted by $D_{1}$), two positive Dirac points at two P's ($D_2$) and eight negative Dirac points along $\Gamma$P and its symmetry equivalents ($D_3$).
More detailed search shows that there is no other band crossing between these two branches.

We remark that the limited experimental data in the literature on this compound cannot fully justify the $J_1$-$J_2$ model (or any spin model), so that the positions of $D_{1,3}$ and even their appearance depend on specifics of the model.
Nonetheless, we emphasize that the high-symmetry point P ($D_2$) is \textit{always} a Dirac point.
This \textit{model independent} Dirac point deserves some detailed analysis given below.
The three screw rotations $R_{x,y,z}$ and $PT$ are elements of the little group at P.
It is straightforward to check that
\begin{eqnarray}
\label{eq:7}\{R_i,R_j\}&=&-2\delta_{ij}.
\end{eqnarray}
$iR_i$'s are hence generators of a Clifford algebra, the simplest choice of which are the Pauli matrices, i.e., $R_i=i\sigma_i$. Since both space inversion and time-reversal commute with $R_i$, $PT$ commutes with $R_i$, so that $R_i$ must be real. But since $i\sigma_{x,z}$ are imaginary, $PT$ excludes this simplest choice. The next choice is that $R_i$ are Dirac matrices, and out of the five generators one can pick $R_x=i\sigma_ys_x$, $R_y=is_y$, $R_z=\pm{i}\sigma_ys_z$, which are real and satisfy Eq.(\ref{eq:7}).
This proof shows that all levels at point P are at least fourfold degenerate.
When U(1)-symmetry is present the two P's in the BZ are found to have the same monopole charge of either $+1$ or $-1$.

\textit{Topological nodal lines in Cu$_3$TeO$_6$}
Due to the less than 180$^\circ$ bond angle of the Cu-O-Cu bond, the DMI generally exists between nearest neighbor spins:
\begin{equation}
H_{DM}=\sum_{\langle{ij}\rangle}D\hat{\mathbf{d}}_{ij}\cdot\mathbf{S}_i\times\mathbf{S}_j,
\end{equation}
where $D$ is the magnitude and $\hat{\mathbf{d}}_{ij}$ is the normal direction of the triangle made from the three atoms in the Cu-O-Cu bond.

The collinear ground state is unstable upon turning on the interaction, but when $D$ is small, there are stable configurations close to the collinear one with spins pointing along a $\langle111\rangle$-direction. In Fig.\ref{fig:3}(a,b), we show the classical ground state configuration for $D/J_1=0.2$ (calculated from quasi-Newton method), and the directions of all spins are given in polar coordinates in the table of Fig.~3(c). This result is fully consistent with the neutron diffraction results \cite{Herak2005}, and in the limit that $D$ is infinitesimally small, it provides a natural explanation for the (collinear) ground state spin orientation along the [111]-direction, which cannot be explained by the Heisenberg model.

\begin{figure}
\includegraphics[width=8.5cm]{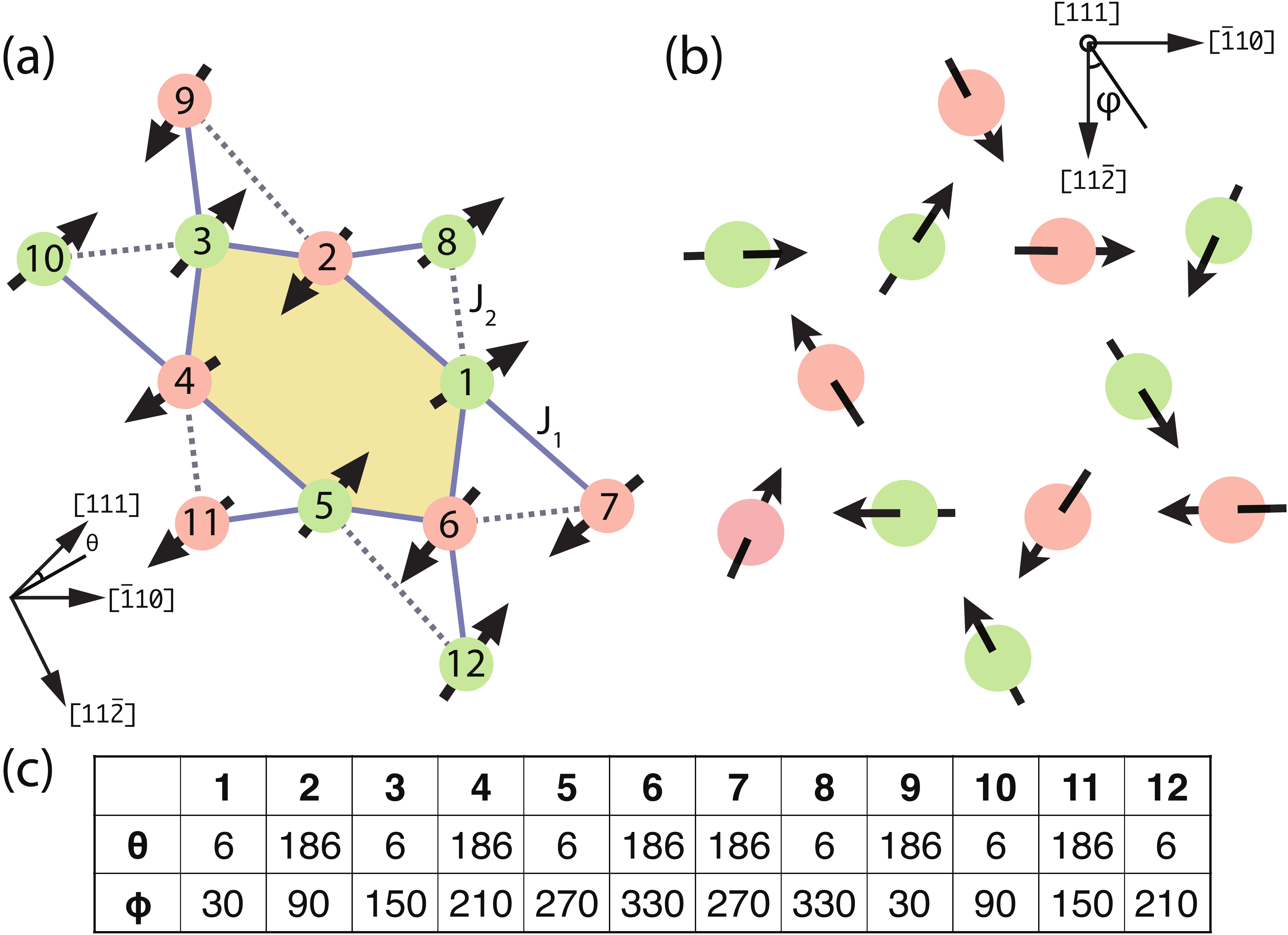}
\caption{\label{fig:3}(a) The classical ground state configuration for $D/J_1$=0.2. Spin-up and -down ions are represented in different colors with nearest-neighbor ($J_1$) and next-nearest-neighbor ($J_2$) interactions indicated. (b) The same configuration projected on the (111)-plane. (c) Table of the exact direction of each polarization, where angle $\theta$ and $\phi$ are defined in the corner panels of (a,b).}
\end{figure}

While the non-collinear ground state breaks many symmetries of the lattice, such as the three screw axes, it preserves $PT$ and threefold rotation along the [111]-direction.
To calculate the spin wave excitations above the non-collinear ground state, one only needs to notice that the spin components in the absolute frame of reference and those in the frame of reference on each site are related by a site-dependent rotation matrix $R_i$. The spin-wave Hamiltonian becomes
\begin{eqnarray}
\nonumber
H'&=&J_1\sum_{\langle{ij}\rangle}R_i\mathbf{S}_i\cdot{R}_j\mathbf{S}_j+J_2\sum_{\langle\langle{ij}\rangle\rangle}R_i\mathbf{S}_i\cdot{R}_j\mathbf{S}_j\\
&+&D\sum_{\langle{ij}\rangle}\hat{\mathbf{d}}_{ij}R_i\mathbf{S}_i\times{R}_j\mathbf{S}_j,
\end{eqnarray}
where the components in $\mathbf{S}_i$ are given in Eq.(\ref{eq:HP1},\ref{eq:HP2}).
Since both the spin interactions and the non-collinear ground state configuration preserve $PT$ and $C_3$, $H'$ also has these symmetries.
The experimental tilting angle is small, implying that DMI may be considered as a perturbation to the original Hamiltonian in Eq.(3). In this case, we can expand $R_i$ in powers of $D$, and collect all terms up to $D^2$ into $\delta{H}\equiv{H'}-H$.
To gain an understanding of how DMI affects the Dirac points, we project $\delta{H}$ onto the subspace spanned by the four degenerate states at P, finding a $k\cdot{p}$ effective Hamiltonian for the spin waves near P. Because of the $C_3$-symmetry, we expect a Dirac point at P either breaks into a ring around the [111]-direction or extends into a straight line along [111], which may be considered the limiting case of an eclipse with a vanishing short axis (see the upper panel of Fig.\ref{fig:bands}(b) for schematics of the two scenarios). In Ref.[\onlinecite{SupMat}], we show that both scenarios may happen, depending on which four degenerate states at P are considered: (i) the Dirac point between the first and the second band (both degenerate) at P becomes a nodal ring and (ii) the Dirac point between the third and the fourth band is stretched into a straight line. In both cases, however, the length of the nodal line is found proportional to $D^2$, and the center of the nodal line is displaced from P by a distance proportional to $D$. Here we only picked Dirac points at P for this analysis; this is because they are the only Dirac point whose existence and position are independent of specifics of the Heisenberg model, and are hence mostly likely to be observed in experiments.

\textit{Discussion}
Finally, we remark on possible experiments that will be able to justify our assumptions and testify to our predictions.
The Dirac points as well as nodal rings in the bulk can be directly measured with inelastic neutron scattering, and they are further expected to exhibit gap-opening behaviors in a magnetic field.
Since magnons of each spin form many Weyl points, there are thermal Hall currents for each spin component. However, because the total Hall current of magnons must vanish due to $PT$, a spin-resolved measurement of the magnon currents is required to observe this effect.
The surface arcs states, however interesting [see Ref.[\onlinecite{SupMat}] for detailed calculation], are difficult to directly observe by inelastic neutron scattering due to the very small sample volume from the surfaces.
One may be able to detect these states using surface-sensitive probes, such as high-resolution electron energy loss spectroscopy, or helium atom energy loss spectroscopy.

\acknowledgements{We wish to thank Ji Feng, Fa Wang, Xuetao Zhu, Ling Lu, and R. Ganesh for discussions and Weiliang Yao for assistance in checking some of our early calculations. The work at Institute of Physics was supported by the National Key Research and Development Program of China under grant No. 2016YFA0302400, by NSFC under grant No. 11674370, 1190020, 11534014, 11334012, by the Ministry of Science and Technology of China 973 program (Grant No. 2015CB921300), and by the Strategic Priority Research Program of CAS (Grant No. XDB07000000). The work at Peking University was supported by the National Natural Science Foundation of China (Grants No. 11374024 and No. 11522429) and Ministry of Science and Technology of China (Grants No. 2015CB921302 and No. 2013CB921903).}

\title{Supplemental Material for ``Dirac and nodal line magnons in three-dimensional antiferromagnets''}
\author{Kangkang Li}
\thanks{These authors contributed equally to this study.}
\affiliation{Beijing National Laboratory for Condensed Matter Physics,
and Institute of Physics, Chinese Academy of Sciences, Beijing 100190, China}
\affiliation{University of Chinese Academy of Sciences, Beijing 100049}
\author{Chenyuan Li}
\thanks{These authors contributed equally to this study.}
\affiliation{International Center for Quantum Materials, School of Physics, Peking University, Beijing 100871, China}
\author{Jiangping Hu}
\affiliation{Beijing National Laboratory for Condensed Matter Physics,
and Institute of Physics, Chinese Academy of Sciences, Beijing 100190, China}
\affiliation{Collaborative Innovation Center of Quantum Matter, Beijing 100871, China}
\author{Yuan Li}
\email{yuan.li@pku.edu.cn}
\affiliation{International Center for Quantum Materials, School of Physics, Peking University, Beijing 100871, China}
\affiliation{Collaborative Innovation Center of Quantum Matter, Beijing 100871, China}
\author{Chen Fang}
\email{cfang@iphy.ac.cn}
\affiliation{Beijing National Laboratory for Condensed Matter Physics,
and Institute of Physics, Chinese Academy of Sciences, Beijing 100190, China}
\maketitle
\onecolumngrid

\section{Appendix A: Spin wave Hamiltonian for Cu$_3$TeO$_6$}
In this supplemental section, we construct the spin wave Hamiltonian for Cu$_3$TeO$_6$ and consequently show that it decouples into two separate sectors, one for each spin quantum number. To illustrate the case, we denote operators for six up spins from 1-6, and the number for each down spin is the same as that for the up spin connected by centro-symmetry. The information of atom positions is from Pearson database\cite{Pearson2}.

Starting from the Heisenberg model Eq.(3), we perform the standard Holstein-Primakoff transformation (Eq.(4,5)) on up and down spins respectively as explained in the main text.
Since the magnetic moments on nearest neighbor atoms are antiparallel, the nearest neighbor spin interaction terms have the general form:
\begin{align}
J_1S_i\cdot S_j=J_1S(a_i^\dagger a_i+b_j^\dagger b_j+a_ib_j+a_i^\dagger b_j^\dagger),
\end{align}
where we assume that $S_i$ pointing upward and $S_j$ downward. The magnetic moments on next-nearest-neighbor atoms are parallel, so the spin interaction terms for up spins have the general form:
\begin{align}
J_2S_i\cdot S_j=J_2S(-a_i^\dagger a_i-a_j^\dagger a_j+a_i^\dagger a_j+a_ia_j^\dagger),
\end{align}
while for down spins $a_i^\dagger, a_i$ are replaced with $b_i^\dagger, b_i$.

After the Fourier transform $a_{\mathbf{q}}=\frac{1}{\sqrt{N_{site}}}\sum_{\mathbf{R}}a_{i\mathbf{R}}e^{-i\mathbf{q}\cdot(\mathbf{R}+\mathbf{d}_i)}$ and $b_{\mathbf{q}}=\frac{1}{\sqrt{N_{site}}}\sum_{\mathbf{R}}b_{i\mathbf{R}}e^{-i\mathbf{q}\cdot(\mathbf{R}+\mathbf{d}_i)}$, the linear spin wave Hamiltonian reads
\begin{align}
H=H_{NN}+H_{NNN}
=\sum_\mathbf{q}\Psi^\dagger(\mathbf{q})H(\mathbf{q})\Psi(\mathbf{q}),
\end{align}
in the basis of $\Psi^\dagger(\mathbf{q})=(
a_{1,\mathbf{q}}^\dagger,a_{2,\mathbf{q}}^\dagger,\cdots,a_{6,\mathbf{q}}^\dagger,b_{1,-\mathbf{q}},b_{2,-\mathbf{q}},
\cdots,b_{6,-\mathbf{q}},
a_{1,-\mathbf{q}},a_{2,-\mathbf{q}},
\cdots ,a_{6,-\mathbf{q}},b^\dag_{1,\mathbf{q}},b^\dag_{2,\mathbf{q}},\cdots,b^\dag_{6,\mathbf{q}})$. Note that this Hamiltonian is block-diagonalized:
\begin{align}
\hat{H}=\sum_\mathbf{q}\Psi^\dagger(\mathbf{q})
\left(
\begin{array}{cc}
H_+(\mathbf{q})&0\\
0&H_-(\mathbf{q})            \\
\end{array}
\right)\Psi(\mathbf{q}),
\end{align}
where $H_+(\mathbf{q})$ and $H_-(\mathbf{q})$ are 12-by-12 Hermitian matrices that satisfy $H_+(k)=H_-^*(k)$ due to U(1) and $PT$ symmetry.

To ``diagonalize'' $H_\pm(\mathbf{q})$, we for example consider the equation of motion of the basis vector of $H_+$, namely, $\Psi_+(\mathbf{q})=(a_{i=1,...,6,\mathbf{q}},b^\dag_{j=1,...,6,-\mathbf{q}})^T$:
\begin{align}
i\dot{a}_{i,\mathbf{q}}=[a_{i,\mathbf{q}},\hat{H}]=\sum_{j=1,\cdots,6}H_+^{i,j}(\mathbf{q})a_{j,\mathbf{q}}+\sum_{j'=1,\cdots,6}H_+^{i,j'+6}(\mathbf{q})b^\dag_{j',-\mathbf{q}},\\
\nonumber
i\dot{b}^\dag_{i,-\mathbf{q}}=[{b}^\dag_{i,-\mathbf{q}},\hat{H}]=-\sum_{j=1,\cdots,6}H_+^{i+6,j}(\mathbf{q})a_{j,\mathbf{q}}-\sum_{j'=1,\cdots,6}H_+^{i+6,j'+6}(\mathbf{q})b^\dag_{j',-\mathbf{q}},
\end{align}
or simply
\begin{align}
E{\Psi_+}(\mathbf{q})=\left(\begin{matrix}
I_6 & 0\\
0 & -I_6\\
\end{matrix}\right)H_+(\mathbf{q})\Psi_+(\mathbf{q}).
\end{align}
\section{Appendix B: Magnon band structures at other parameters}
\label{sec:otherJ2}
\begin{figure}
\includegraphics[width=12cm]{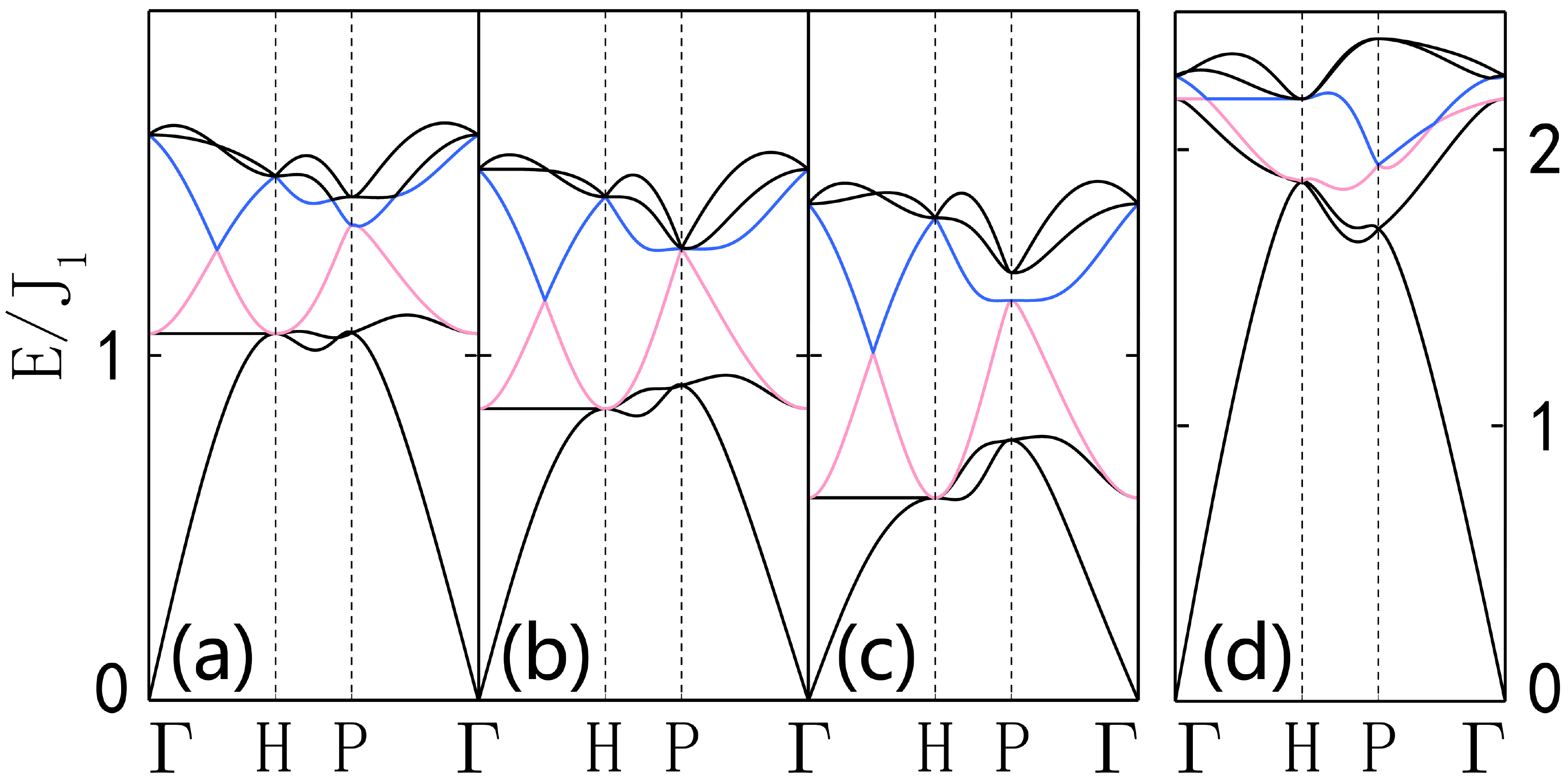}
\caption{The bulk band structure of the spin waves for $J_2/J_1=0.18, 0.23, 0.28, -0.134$, respectively.}
\label{fig:otherJ2}
\end{figure}

At this point, we do not know the relative values between $J_2$ and $J_1$ except that $J_2<J_1/3$ for the stability of the collinear antiferromagetic ground state. In the main text, the value of $J_2=0.134J_1$ was chosen so that the three types of Dirac points have similar energy, optimizing the visibility of the surface arcs. Here, we show the band structures of the spin wave Hamiltonian for different $J_2/J_1=0.18, 0.23, 0.28, -0.134$ respectively in Fig.\ref{fig:otherJ2}.
In Fig.\ref{fig:otherJ2}(a-c) we see that a ``band inversion'' actually happens at $P$ point as one increases $J_2$, and that despite the difference, the Dirac points along $\Gamma{H}$ and those at $P$ remain. In Fig.\ref{fig:otherJ2}(d) we see that even a ferromagnetic $J_2$ cannot alter the main results: all Dirac points $D_{1,2,3}$ still exist along $\Gamma{H}$, at $P$ and along $\Gamma{P}$, while moving to different energies.

\section{Appendix C: $Z_2$-monopole charge of the Dirac point}

In the main text, we claim that each Dirac point protected by $PT$-symmetry and U(1)-symmetry, is a special case of a nodal ring protected by $PT$ when U(1) is broken by perturbation, and that each nodal ring emerging from a Dirac point when U(1)-symmetry is broken must have nontrivial $Z_2$-monopole charge.

Before going to the calculation, let us first comment on the physical consequence of the $Z_2$-monopole charge. All nodal rings so far studied in the literature (with the exception of double-nodal ring) are protected by the Berry phase of $\pi$, along any loop that encircles the nodal line, which is a $Z_2$-index. But nodal lines protected by $PT$ in three dimensions can have a second $Z_2$ invariant, defined on a sphere enclosing the whole loop. When this invariant is zero (trivial), the nodal loop inside can continuously shrink to a point and be gapped; when it is nonzero, the nodal ring inside the sphere can shrink to a point but not be gapped. In fact, suppose one tunes some parameter monotonically, then one can see the nodal ring gradually shrinks to a point, but after that the point grows back into a nodal ring with finite radius. For more information on this $Z_2$-invariant, see Ref.[\onlinecite{Fang2015a2}].

In this work, at each Dirac point, each of the two sectors $H_\pm$ contributes a Weyl point Hamiltonian, and we have $H_+=H_-^\ast$, so the four-by-four $k\cdot{p}$-model near the Dirac point is
\begin{align}
H(\mathbf{q})=q_x\Sigma_{0x}+q_y\Sigma_{zy}+q_z\Sigma_{0z},
\end{align}
where $\Sigma_{\mu\nu=0,x,y,z}=\sigma_\mu\otimes\sigma_\nu$, and the symmetry is represented by
\begin{align}
PT=K\Sigma_{x0}.
\end{align}
We make a basis change by
\begin{align}
U=\left(\begin{matrix}
i & -1\\
-i & -1
\end{matrix}\right)\otimes\sigma_0/\sqrt{2},
\end{align}
after which we have
\begin{align}\label{eq:sup1}
\nonumber
PT\rightarrow{}K\\
H(\mathbf{q})\rightarrow{}q_x\Sigma_{0x}-q_y\Sigma_{yy}+q_z\Sigma_{0z}.
\end{align}
Eq.(\ref{eq:sup1}) is exactly the same as Eq.(5) of Ref.[\onlinecite{Fang2015a2}] without a mass term, where it is shown by explicit calculation to have nontrivial $Z_2$ monopole charge.

\section{Appendix D: topological nodal lines in a four-band $k\cdot p$ model}
We apply the general theory of symmetry-protected nodal line to a specific Dirac point P($D_2$), and accordingly illustrate how the Dirac point evolves into a nodal line when breaking U(1)-symmetry but still preserving $P*T$-symmetry.

To begin with, the four-by-four $k\cdot p$-model near the four degenerate states at P is
\begin{align}
H_{eff}(\mathbf{q})=\sum_{i,j=1}^{4}\bigg ( \langle\psi_i | H_{q_x}^{'}(P)|\psi_j\rangle q_x+\langle\psi_i | H_{q_y}^{'}(P)|\psi_j\rangle q_y+\langle\psi_i | H_{q_z}^{'}(P)|\psi_j\rangle q_z\bigg).
\end{align}

For the crossing of the first and the second band (each doubly degenerate) at P, the effective Hamiltonian without DMI is
\begin{align}
H_{eff_{12}}(\mathbf{q})=0.0284(q_m\Sigma_{0x}+q_n\Sigma_{zy}+q_l\Sigma_{0z}),
\end{align}
where $\mathbf{q_m}=(-0.8140,0.3596,0.4556)$, $\mathbf{q_n}=(0.0553,-0.7328,0.6784)$, $\mathbf{q_l}=\frac{1}{\sqrt{3}}(1,1,1)$,
and the frame of reference of the pseudo-spin space is rotated as
\begin{align}
\frac{1}{2}\Sigma_{\mu x}+\frac{\sqrt{3}}{2}\Sigma_{\mu y}\rightarrow{}\Sigma_{\mu x},\\
-\frac{\sqrt{3}}{2}\Sigma_{\mu x}+\frac{1}{2}\Sigma_{\mu y}\rightarrow{}\Sigma_{\mu y}.
\nonumber
\end{align}

For the crossing of the third and the fourth band (each doubly degenerate) at P, similarly we have the effective Hamiltonian without DMI:
\begin{align}
H_{eff_{34}}(\mathbf{q})=0.0284(q_1\Sigma_{0x}+q_2\Sigma_{zy}-q_3\Sigma_{0z}),
\end{align}
where $\mathbf{q_1}=(-0.3591, 0.4549, -0.8153)$, $\mathbf{q_2}=(-0.7326, 0.6780, 0.0553)$, $\mathbf{q_3}=\frac{1}{\sqrt{3}}(1,1,1)$, 
and we have rotated the frame of reference of the pseudo-spin space as
\begin{align}
\frac{1}{2}\Sigma_{\mu x}-\frac{\sqrt{3}}{2}\Sigma_{\mu y}\rightarrow{}\Sigma_{\mu x},\\
\frac{\sqrt{3}}{2}\Sigma_{\mu x}+\frac{1}{2}\Sigma_{\mu y}\rightarrow{}\Sigma_{\mu y}.
\nonumber
\end{align}

Upon turning on DMI, the stable ground state configurations become non-collinear. Thus, we define a local frame of reference for each spin so that the ordered moment is along +z-direction, and perform the standard Holstein-Primakoff transformation as Eq.(4). If the ordered moment $S_i$ is along $(\Theta,\Phi)$ in the absolute frame of reference, its components are related to those in the local frame of reference $\widetilde{S_i}$ by a site-dependent rotational matrix $R_i$ as
\begin{align}
S_i=R_i\widetilde{S_i},
\end{align}
\begin{align}
R_i=R(\Theta,\Phi)=R_{z}(\Phi)R_{y}(\Theta)=\left(\begin{matrix}
\cos\Phi & -\sin\Phi & 0\\
sin\Phi  & \cos\Phi &0\\
0 & 0 & 1
\end{matrix}\right)\left(\begin{matrix}
\cos\Theta & 0 & \sin\Theta\\
0 & 1 & 0\\
-\sin\Theta & 0 & \cos\Theta
\end{matrix}\right),
\end{align}
which rotates vectors by $\Theta$ about the y-axis first and then $\Phi$ about the z-axis. In the case of the collinear ground state with spins pointing along a $\langle 111\rangle$-direction, $R_{i0}$ is $R\big(\arctan{\sqrt{2}},\frac{\pi}{4}\big)$ for spin up and $R\big(\pi+\arctan{\sqrt{2}},\frac{\pi}{4}\big)$ for spin down, which is actually equivalent with the procedure described in Section I.

Considering that D is small and DMI may thus be considered as a perturbation, we draw support from a new frame in the basis of
\begin{align}
\hat{\mathbf{i}}=\frac{1}{\sqrt{6}}(1,1,-2), \quad \hat{\mathbf{j}}=\frac{1}{\sqrt{2}}(-1,1,0), \quad \hat{\mathbf{k}}=\frac{1}{\sqrt{3}}(1,1,1),
\end{align}
so as to better illustrate the deviation from the original direction. The non-collinear ground state could be described by the polar angle $\theta$ and the azimuthal angle $\phi$ in the new frame, as shown in the main text for a specific parameter set. Then we expand $d\Theta$ and $d\Phi$ in powers of $\theta$:
\begin{align}
d\Theta&=\cos \phi \theta+\frac{\sqrt{2}}{4} \sin^2\phi \theta^2,\\
d\Phi&=\frac{\sqrt{6}}{2}\sin\phi \theta-\frac{\sqrt{3}}{2}\cos\phi\sin\phi\theta^2.
\nonumber
\end{align}
By assuming $\theta=cD$, we can further expand the rotational matrix in powers of D,
\begin{align}
R_i=&R_{i0}+\bigg(\cos\phi\frac{\partial R_i}{\partial \Phi}+\frac{\sqrt{6}}{2}\sin\phi\frac{\partial R_i}{\partial \Theta}\bigg)cD+\bigg(-\frac{\sqrt{3}}{2}\cos\phi\sin\phi\frac{\partial R_i}{\partial \Phi}+\frac{\sqrt{2}}{4} \sin^2\phi\frac{\partial R_i}{\partial \Theta}\nonumber\\&+\frac{1}{2}(\frac{\sqrt{6}}{2}\sin\phi)^2\frac{\partial^2 R_i}{\partial \Phi^2}+\frac{\sqrt{6}}{2}\cos\phi\sin\phi\frac{\partial^2 R_i}{\partial \Phi \partial\Theta}+\frac{1}{2}(\cos \phi)^2\frac{\partial^2 R_i}{\partial \Theta^2}\bigg)c^2D^2.
\end{align}
Following the convention of the main text, $\delta H$ expressed by rotational matrices is
\begin{align}
\delta H&=H'-H\nonumber\\
&=J_1\sum_{\langle ij \rangle}\bigg(R_i\mathbf{S}_i\cdot R_j\mathbf{S}_j-R_{i0}\mathbf{S}_i\cdot R_{j0}\mathbf{S}_j\bigg)+J_2\sum_{\langle\langle ij \rangle\rangle}\bigg(R_i\mathbf{S}_i\cdot R_j\mathbf{S}_j-R_{i0}\mathbf{S}_i\cdot R_{j0}\mathbf{S}_j\bigg)
+D\sum_{\langle ij \rangle}\hat{\mathbf{d}}_{ij}\cdot R_i\mathbf{S}_i\times R_j\mathbf{S}_j.
\end{align}
Therefore, the spin wave Hamiltonian around P is
\begin{align}
H'_{eff}(\mathbf{q})=H_{eff}(\mathbf{q})+\sum_{i,j=1}^{4}\langle\psi_i | \delta H(P)|\psi_j\rangle.
\end{align}

Before discussing specific results of two crossings at P, we remark that terms of $\Sigma_{00}$ and $\Sigma_{0z}$ are trivial since they do not change the shape of dispersion relation, just effecting the energy level as a whole. Still, the latter can change the position of Dirac point or nodal line in momentum space. Thus, such terms like $\Sigma_{00}$ in $\delta H$ are ignored below.

For the crossing of the first and the second band at P, the new effective Hamiltonian with DMI is
\begin{align}
H'_{eff_{12}}(\mathbf{q})=0.0284(q_m\Sigma_{0x}+q_n\Sigma_{zy})+(0.0284q_l+d)\Sigma_{0z}+a\Sigma_{xx},
\end{align}
where
\begin{align}
d&=0.249D-0.295cD^2+0.370c^2D^2,\\
a&=\sqrt{A^2+B^2}=\sqrt{(0.206cD^2-0.115c^2D^2)^2+(-0.385cD^2+0.429c^2D^2)^2},
\nonumber
\end{align}
and the frame of reference of the pseudo-spin space is rotated as
\begin{align}
\frac{A}{\sqrt{A^2+B^2}}\Sigma_{x\nu}+\frac{B}{\sqrt{A^2+B^2}}\Sigma_{y\nu}\rightarrow{}\Sigma_{x\nu},\\
-\frac{B}{\sqrt{A^2+B^2}}\Sigma_{x\nu}+\frac{A}{\sqrt{A^2+B^2}}\Sigma_{y\nu}\rightarrow{}\Sigma_{y\nu}.
\nonumber
\end{align}
If $d'$ and $a'$ are defined as $d'=d/0.0284, a'=a/0.0284$, then the spectrum is given by
\begin{align}
E(\mathbf{q})=\pm0.0284\sqrt{(a'\pm \sqrt{q_m^2+q_n^2})^2+(q_l+d')^2},
\end{align}
which is exactly the same as Eq.(5) of Ref.\cite{Fang2015a2} with a mass term if we replace $q_l$ with $q_l'=q_l+d$. The Dirac point at P breaks into a nodal ring perpendicular to the [111]-direction with its radius proportional to $D^2$ if $a\neq 0$. As $a$ changes from positive to negative, the nodal line on $q_l=-d$ and $\sqrt{q_m^2+q_n^2}=|a'|$ decreases and shrinks to a point at $a=0$, and then increases again.

For the crossing of the third and the fourth band at P, the new effective Hamiltonian with DMI is
\begin{align}
H'_{eff_{34}}(\mathbf{q})=0.0284(q_1\Sigma_{0x}+q_2\Sigma_{zy})+(d-0.0284q_3)\Sigma_{0z}+a\Sigma_{x0}+a\Sigma_{xz},
\end{align}
where
\begin{align}
d&=0.249D-0.150cD^2+0.215c^2D^2,\\
a&=\sqrt{A^2+B^2}=\sqrt{(-0.127cD^2-0.085c^2D^2)^2+(0.282cD^2-0.249c^2D^2)^2}.
\nonumber
\end{align}
Here we have rotated the frame of reference of the pseudo-spin space:
\begin{align}
\frac{A}{\sqrt{A^2+B^2}}\Sigma_{x\nu}+\frac{B}{\sqrt{A^2+B^2}}\Sigma_{y\nu}\rightarrow{}\Sigma_{x\nu},\\
-\frac{B}{\sqrt{A^2+B^2}}\Sigma_{x\nu}+\frac{A}{\sqrt{A^2+B^2}}\Sigma_{y\nu}\rightarrow{}\Sigma_{y\nu}.
\nonumber
\end{align}
If $d'$ and $a'$ are defined as $d'=d/0.0284, a'=a/0.0284$, then the spectrum is given by
\begin{align}
E(\mathbf{q})=0.0284\bigg(a'\pm\sqrt{(a'+q_3-d')^2+q_1^2+q_2^2}\bigg) , 0.0284\bigg(-a'\pm\sqrt{(a'-q_3+d')^2+q_1^2+q_2^2}\bigg).
\end{align}
The band crossing could be found where $d'-|a'|<q_3<d'+|a'|$ and $q_1=q_2=0$. Therefore, the Dirac point is stretched into a short straight nodal line along the [111]-direction with its length proportional to $D^2$ if $a\neq 0$. As $a$ changes from positive to negative, the nodal line decreases and shrinks to a point at $a=0$, and then increases again. Here the nodal line can be seen as the special case of a closed ring.

\section{Appendix E: Topological surface states with ``double-arcs''}

Due to the widely acknowledged bulk-edge correspondence principle, bulk topological states in $d$-dimensions have anomalous surface states that cannot be realized on a $d-1$-dimensional lattice without breaking certain symmetries.
For topological band crossings, the surface bands form a helicoid structure surrounding the projection of the bulk node.
To be specific, if two bulk bands cross each other at a Dirac point, then in the surface Brillouin zone, inside the gap between the two bands, there are surface states, and the dispersion of the surface states form a double-helicoid (two helicoids winding in different directions) centered at the projection of the Dirac point.
As stated in Ref.[\onlinecite{Fang20162}], the equal energy contours of a double-helicoid are two arcs\cite{Wang20122,Xu20152} emanating from the Dirac point projection, ending at the projection of another Dirac point of opposite charge.

\begin{figure}
\includegraphics[width=8.5cm]{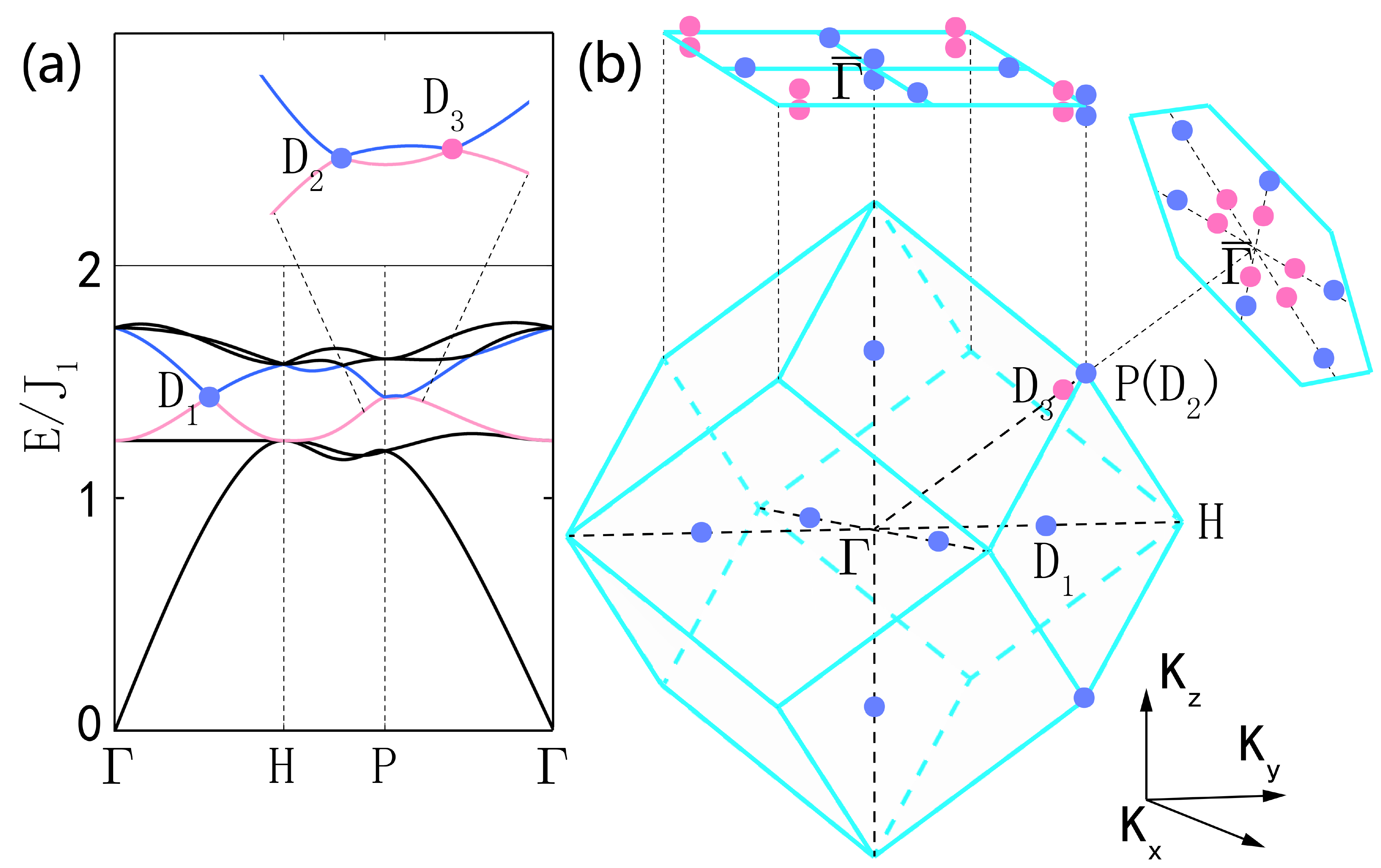}
\caption{(a) A typical band structure of the spin-wave dispersion along high-symmetry lines with $J_2=0.134J_1>0$, where the inset shows a zoomed-in region near P. (b) Positions of all Dirac points in the first Brillouin zone and their projections onto the $(001)$- and $(111)$-surface Brillouin zones. Red and blue colors indicate the monopole charge of +1 and -1, respectively. For clarity, only one of the eight W$_3$ points is displayed in the three-dimensional Brillouin zone in (b).}
\label{fig:bands}
\end{figure}

\begin{figure}
\includegraphics[width=8.5cm]{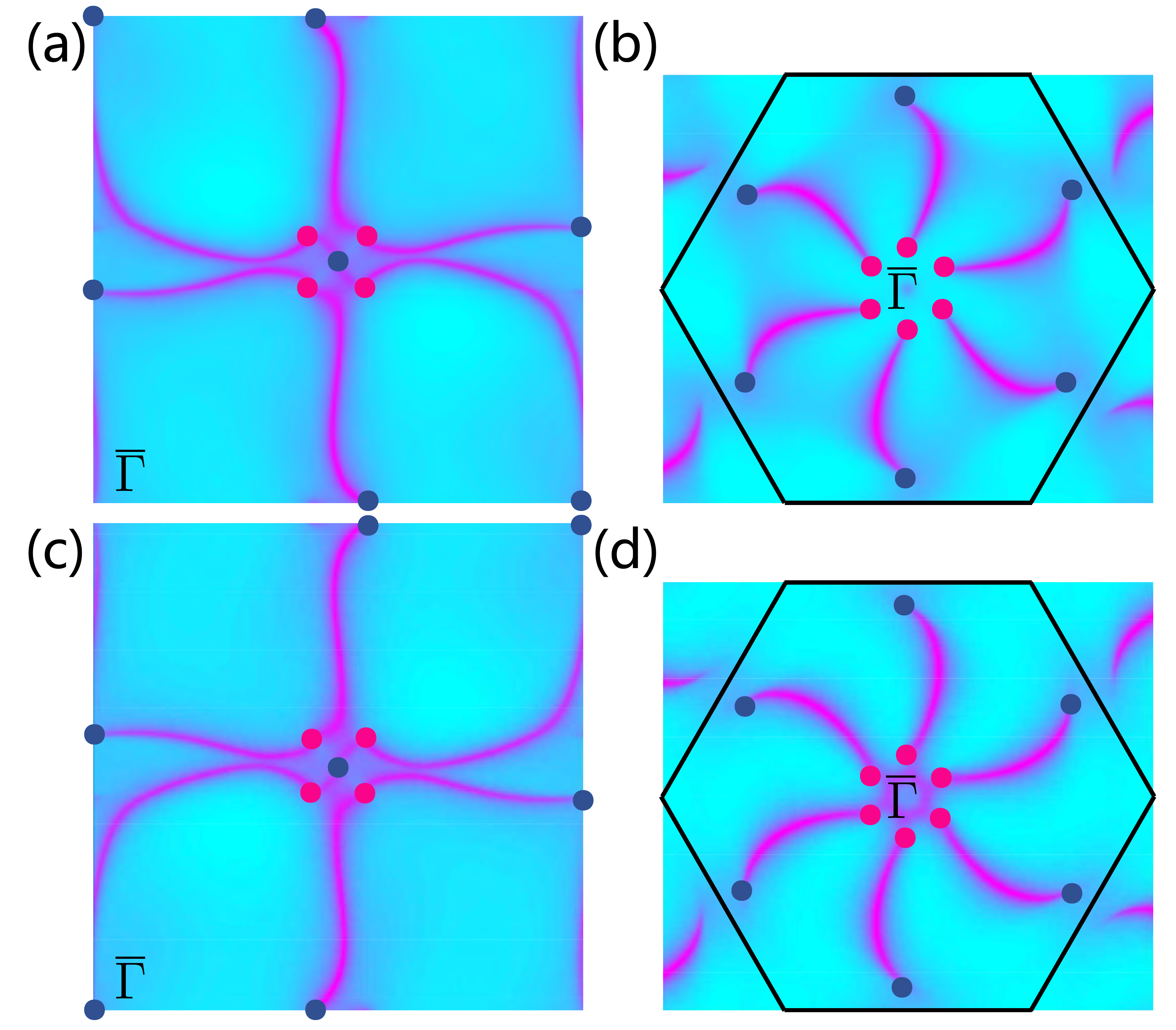}
\caption{The imaginary part of susceptibility $\chi^+(\mathbf{q},\omega=E_0)$ (a, b), $\chi^-(\mathbf{q},\omega=E_0)$ (c, d) on the $(001)$-surface (a, c) and the $(111)$-surface (b, d) at $E_0=1.436J_1$ are shown. The projections of the positively (negatively) charged Dirac points in the surface Brillouin zone are marked in red (deep blue).}
\label{fig:arc}
\end{figure}
We consider two typical open surfaces of the $(001)$-plane and the $(111)$-plane.
For the $(001)$- and $(111)$-surfaces, the surface Brillouin zones are shown in Fig.~\ref{fig:bands}(b) by projecting the bulk Brillouin zone.
The Dirac points are also projected onto the surface Brillouin zone and there are instances when two Dirac points in the bulk Brillouin zone project to the same point on the surface.
We set the parameters to $J_2=0.134J_1$ such that $D_{1,2,3}$ are close in energy.
Then we choose an energy $E_0$ that is the average energy of $D_{1,2,3}$ and plot the equal energy contours in the surface Brillouin zone.

To calculate the dynamic susceptibility on the top surface (or bottom surface), we first do a partial Fourier transform, leaving the coordinate vertical to the open surface in the real space (denoted by $z$), and transforming the two parallel directions into momentum space
\begin{equation}
S^{x,y,z}_i(z,\mathbf{q}_\|)=\frac{1}{\sqrt{N}}\sum_{\mathbf{R}_\|}S^{x,y,z}_i(\mathbf{R})e^{-i\mathbf{q}_{\|}\cdot\mathbf{R}_{\|}},
\end{equation}
where $i$ runs through the twelve spins in one unit cell and $\mathbf{R}$ denotes the lattice points.
The susceptibility on the top surface is defined as
\begin{equation}
\chi^{\pm}(\mathbf{q}_\|,\omega)=\int_0^\infty{dt}e^{i\omega{t}}\langle{S}^\pm(z=0,\mathbf{q}_\|,t)S^\mp(z=0,-\mathbf{q}_\|,t=0)\rangle,
\end{equation}
where $\langle\rangle$ takes the expectation value over the ground state. The susceptibility, after rewriting the spin operators in terms of the spin wave operators, is nothing but the Green's function of the spin wave field on the top surface. These Green's functions for semi-infinite systems are calculated using the iterative method proposed in Ref.[\onlinecite{Lopez19842, Lopez19852}].

$\chi^\pm(\mathbf{q},\omega)$ receives contribution from $H_+$ and $H_-$ respectively. For clarity, the imaginary parts of $\chi^\pm(\mathbf{q},\omega=E_0)$ are separately plotted in Fig.~\ref{fig:arc}, yet in a spin-unpolarized experiment, the two patterns (upper and lower) should be superimposed.
Since the energy dependence of $\chi(\mathbf{q},\omega)$ takes the form $(\omega-E(\mathbf{q})-i\eta)^{-1}$, where $E(\mathbf{q})$ is the energy dispersion of the surface band, the maxima of the imaginary part of $\chi(\mathbf{q},E_0)$ give the equal energy contour of $E(\mathbf{q})$ at $E_0$.
We see that the projection of each positive Dirac point is connected to that of a negative Dirac point by an arc and vice versa; and the projection of two positive Dirac points is connected to the two projections of negative Dirac points and vice versa.
Since the positive and negative Dirac points are set far apart in the Brillouin zone, the arcs connecting their projections are long, a desirable feature for experimental observation.
Additionally, we note that due to the symmetry $PT$, the two opposite surfaces have identical dispersions of the surface states.


\begin{thebibliography}{44}%
\makeatletter
\providecommand \@ifxundefined [1]{%
 \@ifx{#1\undefined}
}%
\providecommand \@ifnum [1]{%
 \ifnum #1\expandafter \@firstoftwo
 \else \expandafter \@secondoftwo
 \fi
}%
\providecommand \@ifx [1]{%
 \ifx #1\expandafter \@firstoftwo
 \else \expandafter \@secondoftwo
 \fi
}%
\providecommand \natexlab [1]{#1}%
\providecommand \enquote  [1]{``#1''}%
\providecommand \bibnamefont  [1]{#1}%
\providecommand \bibfnamefont [1]{#1}%
\providecommand \citenamefont [1]{#1}%
\providecommand \href@noop [0]{\@secondoftwo}%
\providecommand \href [0]{\begingroup \@sanitize@url \@href}%
\providecommand \@href[1]{\@@startlink{#1}\@@href}%
\providecommand \@@href[1]{\endgroup#1\@@endlink}%
\providecommand \@sanitize@url [0]{\catcode `\\12\catcode `\$12\catcode
  `\&12\catcode `\#12\catcode `\^12\catcode `\_12\catcode `\%12\relax}%
\providecommand \@@startlink[1]{}%
\providecommand \@@endlink[0]{}%
\providecommand \url  [0]{\begingroup\@sanitize@url \@url }%
\providecommand \@url [1]{\endgroup\@href {#1}{\urlprefix }}%
\providecommand \urlprefix  [0]{URL }%
\providecommand \Eprint [0]{\href }%
\providecommand \doibase [0]{http://dx.doi.org/}%
\providecommand \selectlanguage [0]{\@gobble}%
\providecommand \bibinfo  [0]{\@secondoftwo}%
\providecommand \bibfield  [0]{\@secondoftwo}%
\providecommand \translation [1]{[#1]}%
\providecommand \BibitemOpen [0]{}%
\providecommand \bibitemStop [0]{}%
\providecommand \bibitemNoStop [0]{.\EOS\space}%
\providecommand \EOS [0]{\spacefactor3000\relax}%
\providecommand \BibitemShut  [1]{\csname bibitem#1\endcsname}%
\let\auto@bib@innerbib\@empty
\bibitem [{\citenamefont {Murakami}(2007)}]{Murakami2007}%
  \BibitemOpen
  \bibfield  {author} {\bibinfo {author} {\bibfnamefont {S.}~\bibnamefont
  {Murakami}},\ }\href@noop {} {\bibfield  {journal} {\bibinfo  {journal} {New
  Journal of Physics}\ }\textbf {\bibinfo {volume} {9}},\ \bibinfo {pages}
  {356} (\bibinfo {year} {2007})}\BibitemShut {NoStop}%
\bibitem [{\citenamefont {Wan}\ \emph {et~al.}(2011)\citenamefont {Wan},
  \citenamefont {Turner}, \citenamefont {Vishwanath},\ and\ \citenamefont
  {Savrasov}}]{Wan2011}%
  \BibitemOpen
  \bibfield  {author} {\bibinfo {author} {\bibfnamefont {X.}~\bibnamefont
  {Wan}}, \bibinfo {author} {\bibfnamefont {A.~M.}\ \bibnamefont {Turner}},
  \bibinfo {author} {\bibfnamefont {A.}~\bibnamefont {Vishwanath}}, \ and\
  \bibinfo {author} {\bibfnamefont {S.~Y.}\ \bibnamefont {Savrasov}},\
  }\href@noop {} {\bibfield  {journal} {\bibinfo  {journal} {Phys. Rev. B}\
  }\textbf {\bibinfo {volume} {{83}}},\ \bibinfo {pages} {205101} (\bibinfo
  {year} {{2011}})}\BibitemShut {NoStop}%
\bibitem [{\citenamefont {Lv}\ \emph {et~al.}(2015)\citenamefont {Lv},
  \citenamefont {Weng}, \citenamefont {Fu}, \citenamefont {Wang}, \citenamefont
  {Miao}, \citenamefont {Ma}, \citenamefont {Richard}, \citenamefont {Huang},
  \citenamefont {Zhao}, \citenamefont {Chen}, \citenamefont {Fang},
  \citenamefont {Dai}, \citenamefont {Qian},\ and\ \citenamefont
  {Ding}}]{Lv2015}%
  \BibitemOpen
  \bibfield  {author} {\bibinfo {author} {\bibfnamefont {B.~Q.}\ \bibnamefont
  {Lv}}, \bibinfo {author} {\bibfnamefont {H.~M.}\ \bibnamefont {Weng}},
  \bibinfo {author} {\bibfnamefont {B.~B.}\ \bibnamefont {Fu}}, \bibinfo
  {author} {\bibfnamefont {X.~P.}\ \bibnamefont {Wang}}, \bibinfo {author}
  {\bibfnamefont {H.}~\bibnamefont {Miao}}, \bibinfo {author} {\bibfnamefont
  {J.}~\bibnamefont {Ma}}, \bibinfo {author} {\bibfnamefont {P.}~\bibnamefont
  {Richard}}, \bibinfo {author} {\bibfnamefont {X.~C.}\ \bibnamefont {Huang}},
  \bibinfo {author} {\bibfnamefont {L.~X.}\ \bibnamefont {Zhao}}, \bibinfo
  {author} {\bibfnamefont {G.~F.}\ \bibnamefont {Chen}}, \bibinfo {author}
  {\bibfnamefont {Z.}~\bibnamefont {Fang}}, \bibinfo {author} {\bibfnamefont
  {X.}~\bibnamefont {Dai}}, \bibinfo {author} {\bibfnamefont {T.}~\bibnamefont
  {Qian}}, \ and\ \bibinfo {author} {\bibfnamefont {H.}~\bibnamefont {Ding}},\
  }\href {\doibase 10.1103/PhysRevX.5.031013} {\bibfield  {journal} {\bibinfo
  {journal} {Phys. Rev. X}\ }\textbf {\bibinfo {volume} {5}},\ \bibinfo {pages}
  {031013} (\bibinfo {year} {2015})}\BibitemShut {NoStop}%
\bibitem [{\citenamefont {Xu}\ \emph {et~al.}(2015)\citenamefont {Xu},
  \citenamefont {Belopolski}, \citenamefont {Alidoust}, \citenamefont
  {Neupane}, \citenamefont {Zhang}, \citenamefont {Sankar}, \citenamefont
  {Huang}, \citenamefont {Lee}, \citenamefont {Chang}, \citenamefont {Wang},
  \citenamefont {Bian}, \citenamefont {Zheng}, \citenamefont {Sanchez},
  \citenamefont {Chou}, \citenamefont {Lin}, \citenamefont {Jia},\ and\
  \citenamefont {Hasan}}]{Xu2015a}%
  \BibitemOpen
  \bibfield  {author} {\bibinfo {author} {\bibfnamefont {S.-Y.}\ \bibnamefont
  {Xu}}, \bibinfo {author} {\bibfnamefont {I.}~\bibnamefont {Belopolski}},
  \bibinfo {author} {\bibfnamefont {N.}~\bibnamefont {Alidoust}}, \bibinfo
  {author} {\bibfnamefont {M.}~\bibnamefont {Neupane}}, \bibinfo {author}
  {\bibfnamefont {C.}~\bibnamefont {Zhang}}, \bibinfo {author} {\bibfnamefont
  {R.}~\bibnamefont {Sankar}}, \bibinfo {author} {\bibfnamefont {S.-M.}\
  \bibnamefont {Huang}}, \bibinfo {author} {\bibfnamefont {C.-C.}\ \bibnamefont
  {Lee}}, \bibinfo {author} {\bibfnamefont {G.}~\bibnamefont {Chang}}, \bibinfo
  {author} {\bibfnamefont {B.}~\bibnamefont {Wang}}, \bibinfo {author}
  {\bibfnamefont {G.}~\bibnamefont {Bian}}, \bibinfo {author} {\bibfnamefont
  {H.}~\bibnamefont {Zheng}}, \bibinfo {author} {\bibfnamefont {D.~S.}\
  \bibnamefont {Sanchez}}, \bibinfo {author} {\bibfnamefont {F.}~\bibnamefont
  {Chou}}, \bibinfo {author} {\bibfnamefont {H.}~\bibnamefont {Lin}}, \bibinfo
  {author} {\bibfnamefont {S.}~\bibnamefont {Jia}}, \ and\ \bibinfo {author}
  {\bibfnamefont {M.~Z.}\ \bibnamefont {Hasan}},\ }\href@noop {} {\bibfield
  {journal} {\bibinfo  {journal} {Science}\ }\textbf {\bibinfo {volume}
  {349}},\ \bibinfo {pages} {613} (\bibinfo {year} {2015})}\BibitemShut
  {NoStop}%
\bibitem [{\citenamefont {Lu}\ \emph {et~al.}(2015)\citenamefont {Lu},
  \citenamefont {Wang}, \citenamefont {Ye}, \citenamefont {Ran}, \citenamefont
  {Fu}, \citenamefont {Joannopoulos},\ and\ \citenamefont {Soljacic}}]{Lu2015}%
  \BibitemOpen
  \bibfield  {author} {\bibinfo {author} {\bibfnamefont {L.}~\bibnamefont
  {Lu}}, \bibinfo {author} {\bibfnamefont {Z.}~\bibnamefont {Wang}}, \bibinfo
  {author} {\bibfnamefont {D.}~\bibnamefont {Ye}}, \bibinfo {author}
  {\bibfnamefont {L.}~\bibnamefont {Ran}}, \bibinfo {author} {\bibfnamefont
  {L.}~\bibnamefont {Fu}}, \bibinfo {author} {\bibfnamefont {J.~D.}\
  \bibnamefont {Joannopoulos}}, \ and\ \bibinfo {author} {\bibfnamefont
  {M.}~\bibnamefont {Soljacic}},\ }\href@noop {} {\bibfield  {journal}
  {\bibinfo  {journal} {Science}\ }\textbf {\bibinfo {volume} {349}},\ \bibinfo
  {pages} {622} (\bibinfo {year} {2015})}\BibitemShut {NoStop}%
\bibitem [{\citenamefont {Burkov}(2016)}]{Burkov2016}%
  \BibitemOpen
  \bibfield  {author} {\bibinfo {author} {\bibfnamefont {A.~A.}\ \bibnamefont
  {Burkov}},\ }\href {http://dx.doi.org/10.1038/nmat4788} {\bibfield  {journal}
  {\bibinfo  {journal} {Nat Mater}\ }\textbf {\bibinfo {volume} {15}},\
  \bibinfo {pages} {1145} (\bibinfo {year} {2016})}\BibitemShut {NoStop}%
\bibitem [{\citenamefont {Haldane}\ and\ \citenamefont
  {Raghu}(2008)}]{Haldane2008}%
  \BibitemOpen
  \bibfield  {author} {\bibinfo {author} {\bibfnamefont {F.~D.~M.}\
  \bibnamefont {Haldane}}\ and\ \bibinfo {author} {\bibfnamefont
  {S.}~\bibnamefont {Raghu}},\ }\href {\doibase 10.1103/PhysRevLett.100.013904}
  {\bibfield  {journal} {\bibinfo  {journal} {Phys. Rev. Lett.}\ }\textbf
  {\bibinfo {volume} {100}},\ \bibinfo {pages} {013904} (\bibinfo {year}
  {2008})}\BibitemShut {NoStop}%
\bibitem [{\citenamefont {Lu}\ \emph {et~al.}(2014)\citenamefont {Lu},
  \citenamefont {Joannopoulos},\ and\ \citenamefont {Soljacic}}]{Lu2014}%
  \BibitemOpen
  \bibfield  {author} {\bibinfo {author} {\bibfnamefont {L.}~\bibnamefont
  {Lu}}, \bibinfo {author} {\bibfnamefont {J.~D.}\ \bibnamefont
  {Joannopoulos}}, \ and\ \bibinfo {author} {\bibfnamefont {M.}~\bibnamefont
  {Soljacic}},\ }\href@noop {} {\bibfield  {journal} {\bibinfo  {journal} {Nat
  Photon}\ }\textbf {\bibinfo {volume} {8}},\ \bibinfo {pages} {821} (\bibinfo
  {year} {2014})}\BibitemShut {NoStop}%
\bibitem [{\citenamefont {Huber}(2016)}]{Huber2016}%
  \BibitemOpen
  \bibfield  {author} {\bibinfo {author} {\bibfnamefont {S.~D.}\ \bibnamefont
  {Huber}},\ }\href {http://dx.doi.org/10.1038/nphys3801} {\bibfield  {journal}
  {\bibinfo  {journal} {Nat Phys}\ }\textbf {\bibinfo {volume} {12}},\ \bibinfo
  {pages} {621} (\bibinfo {year} {2016})}\BibitemShut {NoStop}%
\bibitem [{\citenamefont {Lu}\ \emph {et~al.}(2013)\citenamefont {Lu},
  \citenamefont {Fu}, \citenamefont {Joannopoulos},\ and\ \citenamefont
  {Soljacic}}]{Lu2013}%
  \BibitemOpen
  \bibfield  {author} {\bibinfo {author} {\bibfnamefont {L.}~\bibnamefont
  {Lu}}, \bibinfo {author} {\bibfnamefont {L.}~\bibnamefont {Fu}}, \bibinfo
  {author} {\bibfnamefont {J.~D.}\ \bibnamefont {Joannopoulos}}, \ and\
  \bibinfo {author} {\bibfnamefont {M.}~\bibnamefont {Soljacic}},\ }\href@noop
  {} {\bibfield  {journal} {\bibinfo  {journal} {Nature Photonics}\ } (\bibinfo
  {year} {2013})}\BibitemShut {NoStop}%
\bibitem [{\citenamefont {Wang}\ \emph {et~al.}(2016)\citenamefont {Wang},
  \citenamefont {Jian},\ and\ \citenamefont {Yao}}]{Wang2016}%
  \BibitemOpen
  \bibfield  {author} {\bibinfo {author} {\bibfnamefont {L.}~\bibnamefont
  {Wang}}, \bibinfo {author} {\bibfnamefont {S.-K.}\ \bibnamefont {Jian}}, \
  and\ \bibinfo {author} {\bibfnamefont {H.}~\bibnamefont {Yao}},\ }\href
  {\doibase 10.1103/PhysRevA.93.061801} {\bibfield  {journal} {\bibinfo
  {journal} {Phys. Rev. A}\ }\textbf {\bibinfo {volume} {93}},\ \bibinfo
  {pages} {061801} (\bibinfo {year} {2016})}\BibitemShut {NoStop}%
\bibitem [{\citenamefont {Stenull}\ \emph {et~al.}(2016)\citenamefont
  {Stenull}, \citenamefont {Kane},\ and\ \citenamefont
  {Lubensky}}]{Stenull2016}%
  \BibitemOpen
  \bibfield  {author} {\bibinfo {author} {\bibfnamefont {O.}~\bibnamefont
  {Stenull}}, \bibinfo {author} {\bibfnamefont {C.~L.}\ \bibnamefont {Kane}}, \
  and\ \bibinfo {author} {\bibfnamefont {T.~C.}\ \bibnamefont {Lubensky}},\
  }\href {\doibase 10.1103/PhysRevLett.117.068001} {\bibfield  {journal}
  {\bibinfo  {journal} {Phys. Rev. Lett.}\ }\textbf {\bibinfo {volume} {117}},\
  \bibinfo {pages} {068001} (\bibinfo {year} {2016})}\BibitemShut {NoStop}%
\bibitem [{\citenamefont {Li}\ \emph {et~al.}(2016)\citenamefont {Li},
  \citenamefont {Li}, \citenamefont {Kim}, \citenamefont {Balents},
  \citenamefont {Yu},\ and\ \citenamefont {Chen}}]{Li2016}%
  \BibitemOpen
  \bibfield  {author} {\bibinfo {author} {\bibfnamefont {F.-Y.}\ \bibnamefont
  {Li}}, \bibinfo {author} {\bibfnamefont {Y.-D.}\ \bibnamefont {Li}}, \bibinfo
  {author} {\bibfnamefont {Y.~B.}\ \bibnamefont {Kim}}, \bibinfo {author}
  {\bibfnamefont {L.}~\bibnamefont {Balents}}, \bibinfo {author} {\bibfnamefont
  {Y.}~\bibnamefont {Yu}}, \ and\ \bibinfo {author} {\bibfnamefont
  {G.}~\bibnamefont {Chen}},\ }\href {http://dx.doi.org/10.1038/ncomms12691}
  {\bibfield  {journal} {\bibinfo  {journal} {Nature Communications}\ }\textbf
  {\bibinfo {volume} {7}},\ \bibinfo {pages} {12691 EP } (\bibinfo {year}
  {2016})}\BibitemShut {NoStop}%
\bibitem [{\citenamefont {Fransson}\ \emph {et~al.}(2016)\citenamefont
  {Fransson}, \citenamefont {Black-Schaffer},\ and\ \citenamefont
  {Balatsky}}]{Fransson2016}%
  \BibitemOpen
  \bibfield  {author} {\bibinfo {author} {\bibfnamefont {J.}~\bibnamefont
  {Fransson}}, \bibinfo {author} {\bibfnamefont {A.~M.}\ \bibnamefont
  {Black-Schaffer}}, \ and\ \bibinfo {author} {\bibfnamefont {A.~V.}\
  \bibnamefont {Balatsky}},\ }\href {\doibase 10.1103/PhysRevB.94.075401}
  {\bibfield  {journal} {\bibinfo  {journal} {Phys. Rev. B}\ }\textbf {\bibinfo
  {volume} {94}},\ \bibinfo {pages} {075401} (\bibinfo {year}
  {2016})}\BibitemShut {NoStop}%
\bibitem [{\citenamefont {Owerre}(2016{\natexlab{a}})}]{Owerre2016a}%
  \BibitemOpen
  \bibfield  {author} {\bibinfo {author} {\bibfnamefont {S.~A.}\ \bibnamefont
  {Owerre}},\ }\href {http://stacks.iop.org/0953-8984/28/i=38/a=386001}
  {\bibfield  {journal} {\bibinfo  {journal} {Journal of Physics: Condensed
  Matter}\ }\textbf {\bibinfo {volume} {28}},\ \bibinfo {pages} {386001}
  (\bibinfo {year} {2016}{\natexlab{a}})}\BibitemShut {NoStop}%
\bibitem [{\citenamefont {Owerre}(2016{\natexlab{b}})}]{Owerre2016}%
  \BibitemOpen
  \bibfield  {author} {\bibinfo {author} {\bibfnamefont {S.~A.}\ \bibnamefont
  {Owerre}},\ }\href@noop {} {\bibfield  {journal} {\bibinfo  {journal}
  {arXiv:1610.08869}\ } (\bibinfo {year} {2016}{\natexlab{b}})}\BibitemShut
  {NoStop}%
\bibitem [{\citenamefont {Young}\ \emph {et~al.}(2012)\citenamefont {Young},
  \citenamefont {Zaheer}, \citenamefont {Teo}, \citenamefont {Kane},
  \citenamefont {Mele},\ and\ \citenamefont {Rappe}}]{Young2012}%
  \BibitemOpen
  \bibfield  {author} {\bibinfo {author} {\bibfnamefont {S.~M.}\ \bibnamefont
  {Young}}, \bibinfo {author} {\bibfnamefont {S.}~\bibnamefont {Zaheer}},
  \bibinfo {author} {\bibfnamefont {J.~C.~Y.}\ \bibnamefont {Teo}}, \bibinfo
  {author} {\bibfnamefont {C.~L.}\ \bibnamefont {Kane}}, \bibinfo {author}
  {\bibfnamefont {E.~J.}\ \bibnamefont {Mele}}, \ and\ \bibinfo {author}
  {\bibfnamefont {A.~M.}\ \bibnamefont {Rappe}},\ }\href {\doibase
  10.1103/PhysRevLett.108.140405} {\bibfield  {journal} {\bibinfo  {journal}
  {Phys. Rev. Lett.}\ }\textbf {\bibinfo {volume} {108}},\ \bibinfo {pages}
  {140405} (\bibinfo {year} {2012})}\BibitemShut {NoStop}%
\bibitem [{\citenamefont {Wang}\ \emph {et~al.}(2012)\citenamefont {Wang},
  \citenamefont {Sun}, \citenamefont {Chen}, \citenamefont {Franchini},
  \citenamefont {Xu}, \citenamefont {Weng}, \citenamefont {Dai},\ and\
  \citenamefont {Fang}}]{Wang2012}%
  \BibitemOpen
  \bibfield  {author} {\bibinfo {author} {\bibfnamefont {Z.}~\bibnamefont
  {Wang}}, \bibinfo {author} {\bibfnamefont {Y.}~\bibnamefont {Sun}}, \bibinfo
  {author} {\bibfnamefont {X.-Q.}\ \bibnamefont {Chen}}, \bibinfo {author}
  {\bibfnamefont {C.}~\bibnamefont {Franchini}}, \bibinfo {author}
  {\bibfnamefont {G.}~\bibnamefont {Xu}}, \bibinfo {author} {\bibfnamefont
  {H.}~\bibnamefont {Weng}}, \bibinfo {author} {\bibfnamefont {X.}~\bibnamefont
  {Dai}}, \ and\ \bibinfo {author} {\bibfnamefont {Z.}~\bibnamefont {Fang}},\
  }\href {\doibase {10.1103/PhysRevB.85.195320}} {\bibfield  {journal}
  {\bibinfo  {journal} {Phys. Rev. B}\ }\textbf {\bibinfo {volume} {{85}}},\
  \bibinfo {pages} {195320} (\bibinfo {year} {{2012}})}\BibitemShut {NoStop}%
\bibitem [{\citenamefont {Wang}\ \emph {et~al.}(2013)\citenamefont {Wang},
  \citenamefont {Weng}, \citenamefont {Wu}, \citenamefont {Dai},\ and\
  \citenamefont {Fang}}]{Wang2013}%
  \BibitemOpen
  \bibfield  {author} {\bibinfo {author} {\bibfnamefont {Z.}~\bibnamefont
  {Wang}}, \bibinfo {author} {\bibfnamefont {H.}~\bibnamefont {Weng}}, \bibinfo
  {author} {\bibfnamefont {Q.}~\bibnamefont {Wu}}, \bibinfo {author}
  {\bibfnamefont {X.}~\bibnamefont {Dai}}, \ and\ \bibinfo {author}
  {\bibfnamefont {Z.}~\bibnamefont {Fang}},\ }\href {\doibase
  10.1103/PhysRevB.88.125427} {\bibfield  {journal} {\bibinfo  {journal} {Phys.
  Rev. B}\ }\textbf {\bibinfo {volume} {88}},\ \bibinfo {pages} {125427}
  (\bibinfo {year} {2013})}\BibitemShut {NoStop}%
\bibitem [{\citenamefont {Yang}\ and\ \citenamefont
  {Nagaosa}(2014)}]{Yang2014}%
  \BibitemOpen
  \bibfield  {author} {\bibinfo {author} {\bibfnamefont {B.-J.}\ \bibnamefont
  {Yang}}\ and\ \bibinfo {author} {\bibfnamefont {N.}~\bibnamefont {Nagaosa}},\
  }\href {http://dx.doi.org/10.1038/ncomms5898} {\bibfield  {journal} {\bibinfo
   {journal} {Nature Communications}\ }\textbf {\bibinfo {volume} {5}},\
  \bibinfo {pages} {4898 EP } (\bibinfo {year} {2014})}\BibitemShut {NoStop}%
\bibitem [{\citenamefont {Fang}\ \emph
  {et~al.}(2016{\natexlab{a}})\citenamefont {Fang}, \citenamefont {Lu},
  \citenamefont {Liu},\ and\ \citenamefont {Fu}}]{Fang2016}%
  \BibitemOpen
  \bibfield  {author} {\bibinfo {author} {\bibfnamefont {C.}~\bibnamefont
  {Fang}}, \bibinfo {author} {\bibfnamefont {L.}~\bibnamefont {Lu}}, \bibinfo
  {author} {\bibfnamefont {J.}~\bibnamefont {Liu}}, \ and\ \bibinfo {author}
  {\bibfnamefont {L.}~\bibnamefont {Fu}},\ }\href
  {http://dx.doi.org/10.1038/nphys3782} {\bibfield  {journal} {\bibinfo
  {journal} {Nat Phys}\ }\textbf {\bibinfo {volume} {12}},\ \bibinfo {pages}
  {936} (\bibinfo {year} {2016}{\natexlab{a}})}\BibitemShut {NoStop}%
\bibitem [{\citenamefont {Bradlyn}\ \emph {et~al.}(2016)\citenamefont
  {Bradlyn}, \citenamefont {Cano}, \citenamefont {Wang}, \citenamefont
  {Vergniory}, \citenamefont {Felser}, \citenamefont {Cava},\ and\
  \citenamefont {Bernevig}}]{Bradlyn2016}%
  \BibitemOpen
  \bibfield  {author} {\bibinfo {author} {\bibfnamefont {B.}~\bibnamefont
  {Bradlyn}}, \bibinfo {author} {\bibfnamefont {J.}~\bibnamefont {Cano}},
  \bibinfo {author} {\bibfnamefont {Z.}~\bibnamefont {Wang}}, \bibinfo {author}
  {\bibfnamefont {M.~G.}\ \bibnamefont {Vergniory}}, \bibinfo {author}
  {\bibfnamefont {C.}~\bibnamefont {Felser}}, \bibinfo {author} {\bibfnamefont
  {R.~J.}\ \bibnamefont {Cava}}, \ and\ \bibinfo {author} {\bibfnamefont
  {B.~A.}\ \bibnamefont {Bernevig}},\ }\href {\doibase 10.1126/science.aaf5037}
  {\bibfield  {journal} {\bibinfo  {journal} {Science}\ }\textbf {\bibinfo
  {volume} {353}} (\bibinfo {year} {2016}),\ 10.1126/science.aaf5037},\ \Eprint
  {http://arxiv.org/abs/http://science.sciencemag.org/content/353/6299/aaf5037.full.pdf}
  {http://science.sciencemag.org/content/353/6299/aaf5037.full.pdf}\BibitemShut {NoStop}%
\bibitem [{\citenamefont {Weng}\ \emph {et~al.}(2016)\citenamefont {Weng},
  \citenamefont {Fang}, \citenamefont {Fang},\ and\ \citenamefont
  {Dai}}]{Weng2016}%
  \BibitemOpen
  \bibfield  {author} {\bibinfo {author} {\bibfnamefont {H.}~\bibnamefont
  {Weng}}, \bibinfo {author} {\bibfnamefont {C.}~\bibnamefont {Fang}}, \bibinfo
  {author} {\bibfnamefont {Z.}~\bibnamefont {Fang}}, \ and\ \bibinfo {author}
  {\bibfnamefont {X.}~\bibnamefont {Dai}},\ }\href {\doibase
  10.1103/PhysRevB.93.241202} {\bibfield  {journal} {\bibinfo  {journal} {Phys.
  Rev. B}\ }\textbf {\bibinfo {volume} {93}},\ \bibinfo {pages} {241202}
  (\bibinfo {year} {2016})}\BibitemShut {NoStop}%
\bibitem [{\citenamefont {Zhu}\ \emph {et~al.}(2016)\citenamefont {Zhu},
  \citenamefont {Winkler}, \citenamefont {Wu}, \citenamefont {Li},\ and\
  \citenamefont {Soluyanov}}]{Zhu2016}%
  \BibitemOpen
  \bibfield  {author} {\bibinfo {author} {\bibfnamefont {Z.}~\bibnamefont
  {Zhu}}, \bibinfo {author} {\bibfnamefont {G.~W.}\ \bibnamefont {Winkler}},
  \bibinfo {author} {\bibfnamefont {Q.}~\bibnamefont {Wu}}, \bibinfo {author}
  {\bibfnamefont {J.}~\bibnamefont {Li}}, \ and\ \bibinfo {author}
  {\bibfnamefont {A.~A.}\ \bibnamefont {Soluyanov}},\ }\href {\doibase
  10.1103/PhysRevX.6.031003} {\bibfield  {journal} {\bibinfo  {journal} {Phys.
  Rev. X}\ }\textbf {\bibinfo {volume} {6}},\ \bibinfo {pages} {031003}
  (\bibinfo {year} {2016})}\BibitemShut {NoStop}%
\bibitem [{\citenamefont {Chang}\ \emph {et~al.}(2016)\citenamefont {Chang},
  \citenamefont {Xu}, \citenamefont {Huang}, \citenamefont {Sanchez},
  \citenamefont {Hsu}, \citenamefont {Bian}, \citenamefont {Yu}, \citenamefont
  {Belopolski}, \citenamefont {Alidoust}, \citenamefont {Zheng}, \citenamefont
  {Chang}, \citenamefont {Jeng}, \citenamefont {Yang}, \citenamefont {Neupert},
  \citenamefont {Lin},\ and\ \citenamefont {Hasan}}]{Chang2016}%
  \BibitemOpen
  \bibfield  {author} {\bibinfo {author} {\bibfnamefont {G.}~\bibnamefont
  {Chang}}, \bibinfo {author} {\bibfnamefont {S.-Y.}\ \bibnamefont {Xu}},
  \bibinfo {author} {\bibfnamefont {S.-M.}\ \bibnamefont {Huang}}, \bibinfo
  {author} {\bibfnamefont {D.~S.}\ \bibnamefont {Sanchez}}, \bibinfo {author}
  {\bibfnamefont {C.-H.}\ \bibnamefont {Hsu}}, \bibinfo {author} {\bibfnamefont
  {G.}~\bibnamefont {Bian}}, \bibinfo {author} {\bibfnamefont {Z.-M.}\
  \bibnamefont {Yu}}, \bibinfo {author} {\bibfnamefont {I.}~\bibnamefont
  {Belopolski}}, \bibinfo {author} {\bibfnamefont {N.}~\bibnamefont
  {Alidoust}}, \bibinfo {author} {\bibfnamefont {H.}~\bibnamefont {Zheng}},
  \bibinfo {author} {\bibfnamefont {T.-R.}\ \bibnamefont {Chang}}, \bibinfo
  {author} {\bibfnamefont {H.-T.}\ \bibnamefont {Jeng}}, \bibinfo {author}
  {\bibfnamefont {S.~A.}\ \bibnamefont {Yang}}, \bibinfo {author}
  {\bibfnamefont {T.}~\bibnamefont {Neupert}}, \bibinfo {author} {\bibfnamefont
  {H.}~\bibnamefont {Lin}}, \ and\ \bibinfo {author} {\bibfnamefont {M.~Z.}\
  \bibnamefont {Hasan}},\ }\href@noop {} {\bibfield  {journal} {\bibinfo
  {journal} {arXiv:1605.06831v1}\ } (\bibinfo {year} {2016})}\BibitemShut
  {NoStop}%
\bibitem [{\citenamefont {Wieder}\ \emph {et~al.}(2016)\citenamefont {Wieder},
  \citenamefont {Kim}, \citenamefont {Rappe},\ and\ \citenamefont
  {Kane}}]{Wieder2016}%
  \BibitemOpen
  \bibfield  {author} {\bibinfo {author} {\bibfnamefont {B.~J.}\ \bibnamefont
  {Wieder}}, \bibinfo {author} {\bibfnamefont {Y.}~\bibnamefont {Kim}},
  \bibinfo {author} {\bibfnamefont {A.~M.}\ \bibnamefont {Rappe}}, \ and\
  \bibinfo {author} {\bibfnamefont {C.~L.}\ \bibnamefont {Kane}},\ }\href
  {\doibase 10.1103/PhysRevLett.116.186402} {\bibfield  {journal} {\bibinfo
  {journal} {Phys. Rev. Lett.}\ }\textbf {\bibinfo {volume} {116}},\ \bibinfo
  {pages} {186402} (\bibinfo {year} {2016})}\BibitemShut {NoStop}%
\bibitem [{\citenamefont {Yang}\ \emph {et~al.}(2017)\citenamefont {Yang},
  \citenamefont {Bojesen}, \citenamefont {Morimoto},\ and\ \citenamefont
  {Furusaki}}]{Yang2017}%
  \BibitemOpen
  \bibfield  {author} {\bibinfo {author} {\bibfnamefont {B.-J.}\ \bibnamefont
  {Yang}}, \bibinfo {author} {\bibfnamefont {T.~A.}\ \bibnamefont {Bojesen}},
  \bibinfo {author} {\bibfnamefont {T.}~\bibnamefont {Morimoto}}, \ and\
  \bibinfo {author} {\bibfnamefont {A.}~\bibnamefont {Furusaki}},\ }\href
  {\doibase 10.1103/PhysRevB.95.075135} {\bibfield  {journal} {\bibinfo
  {journal} {Phys. Rev. B}\ }\textbf {\bibinfo {volume} {95}},\ \bibinfo
  {pages} {075135} (\bibinfo {year} {2017})}\BibitemShut {NoStop}%
\bibitem [{\citenamefont {Young}\ and\ \citenamefont
  {Wieder}(2016)}]{Wieder2017}%
  \BibitemOpen
  \bibfield  {author} {\bibinfo {author} {\bibfnamefont {S.~M.}\ \bibnamefont
  {Young}}\ and\ \bibinfo {author} {\bibfnamefont {B.~J.}\ \bibnamefont
  {Wieder}},\ }\href@noop {} {\bibfield  {journal} {\bibinfo  {journal}
  {arXiv:1609.06738}\ } (\bibinfo {year} {2016})}\BibitemShut {NoStop}%
\bibitem [{\citenamefont {Watanabe}\ \emph {et~al.}(2016)\citenamefont
  {Watanabe}, \citenamefont {Po}, \citenamefont {Zaletel},\ and\ \citenamefont
  {Vishwanath}}]{Watanabe2016}%
  \BibitemOpen
  \bibfield  {author} {\bibinfo {author} {\bibfnamefont {H.}~\bibnamefont
  {Watanabe}}, \bibinfo {author} {\bibfnamefont {H.~C.}\ \bibnamefont {Po}},
  \bibinfo {author} {\bibfnamefont {M.~P.}\ \bibnamefont {Zaletel}}, \ and\
  \bibinfo {author} {\bibfnamefont {A.}~\bibnamefont {Vishwanath}},\ }\href
  {\doibase 10.1103/PhysRevLett.117.096404} {\bibfield  {journal} {\bibinfo
  {journal} {Phys. Rev. Lett.}\ }\textbf {\bibinfo {volume} {117}},\ \bibinfo
  {pages} {096404} (\bibinfo {year} {2016})}\BibitemShut {NoStop}%
\bibitem [{\citenamefont {BRADLEY}\ and\ \citenamefont
  {DAVIES}(1968)}]{Bradley1968}%
  \BibitemOpen
  \bibfield  {author} {\bibinfo {author} {\bibfnamefont {C.~J.}\ \bibnamefont
  {BRADLEY}}\ and\ \bibinfo {author} {\bibfnamefont {B.~L.}\ \bibnamefont
  {DAVIES}},\ }\href {\doibase 10.1103/RevModPhys.40.359} {\bibfield  {journal}
  {\bibinfo  {journal} {Rev. Mod. Phys.}\ }\textbf {\bibinfo {volume} {40}},\
  \bibinfo {pages} {359} (\bibinfo {year} {1968})}\BibitemShut {NoStop}%
\bibitem [{\citenamefont {Dzyaloshinsky}(1958)}]{Dzyaloshinsky1958}%
  \BibitemOpen
  \bibfield  {author} {\bibinfo {author} {\bibfnamefont {I.}~\bibnamefont
  {Dzyaloshinsky}},\ }\href@noop {} {\bibfield  {journal} {\bibinfo  {journal}
  {Journal of Physics and Chemistry of Solids}\ }\textbf {\bibinfo {volume}
  {4}},\ \bibinfo {pages} {241} (\bibinfo {year} {1958})}\BibitemShut {NoStop}%
\bibitem [{\citenamefont {Moriya}(1960)}]{Moriya1960}%
  \BibitemOpen
  \bibfield  {author} {\bibinfo {author} {\bibfnamefont {T.}~\bibnamefont
  {Moriya}},\ }\href {\doibase 10.1103/PhysRev.120.91} {\bibfield  {journal}
  {\bibinfo  {journal} {Phys. Rev.}\ }\textbf {\bibinfo {volume} {120}},\
  \bibinfo {pages} {91} (\bibinfo {year} {1960})}\BibitemShut {NoStop}%
\bibitem [{\citenamefont {Fang}\ \emph
  {et~al.}(2016{\natexlab{b}})\citenamefont {Fang}, \citenamefont {Weng},
  \citenamefont {Dai},\ and\ \citenamefont {Fang}}]{Fang2016a}%
  \BibitemOpen
  \bibfield  {author} {\bibinfo {author} {\bibfnamefont {C.}~\bibnamefont
  {Fang}}, \bibinfo {author} {\bibfnamefont {H.}~\bibnamefont {Weng}}, \bibinfo
  {author} {\bibfnamefont {X.}~\bibnamefont {Dai}}, \ and\ \bibinfo {author}
  {\bibfnamefont {Z.}~\bibnamefont {Fang}},\ }\href
  {http://stacks.iop.org/1674-1056/25/i=11/a=117106} {\bibfield  {journal}
  {\bibinfo  {journal} {Chinese Physics B}\ }\textbf {\bibinfo {volume} {25}},\
  \bibinfo {pages} {117106} (\bibinfo {year} {2016}{\natexlab{b}})}\BibitemShut
  {NoStop}%
\bibitem [{\citenamefont {Fang}\ \emph {et~al.}(2015)\citenamefont {Fang},
  \citenamefont {Chen}, \citenamefont {Kee},\ and\ \citenamefont
  {Fu}}]{Fang2015a}%
  \BibitemOpen
  \bibfield  {author} {\bibinfo {author} {\bibfnamefont {C.}~\bibnamefont
  {Fang}}, \bibinfo {author} {\bibfnamefont {Y.}~\bibnamefont {Chen}}, \bibinfo
  {author} {\bibfnamefont {H.-Y.}\ \bibnamefont {Kee}}, \ and\ \bibinfo
  {author} {\bibfnamefont {L.}~\bibnamefont {Fu}},\ }\href {\doibase
  10.1103/PhysRevB.92.081201} {\bibfield  {journal} {\bibinfo  {journal} {Phys.
  Rev. B}\ }\textbf {\bibinfo {volume} {92}},\ \bibinfo {pages} {081201}
  (\bibinfo {year} {2015})}\BibitemShut {NoStop}%
\bibitem [{\citenamefont {Burkov}\ \emph {et~al.}(2011)\citenamefont {Burkov},
  \citenamefont {Hook},\ and\ \citenamefont {Balents}}]{Burkov2011}%
  \BibitemOpen
  \bibfield  {author} {\bibinfo {author} {\bibfnamefont {A.~A.}\ \bibnamefont
  {Burkov}}, \bibinfo {author} {\bibfnamefont {M.~D.}\ \bibnamefont {Hook}}, \
  and\ \bibinfo {author} {\bibfnamefont {L.}~\bibnamefont {Balents}},\
  }\href@noop {} {\bibfield  {journal} {\bibinfo  {journal} {Phys. Rev. B}\
  }\textbf {\bibinfo {volume} {{84}}},\ \bibinfo {pages} {235126} (\bibinfo
  {year} {{2011}})}\BibitemShut {NoStop}%
\bibitem [{\citenamefont {Xu}\ \emph {et~al.}(2011)\citenamefont {Xu},
  \citenamefont {Weng}, \citenamefont {Wang}, \citenamefont {Dai},\ and\
  \citenamefont {Fang}}]{Xu2011}%
  \BibitemOpen
  \bibfield  {author} {\bibinfo {author} {\bibfnamefont {G.}~\bibnamefont
  {Xu}}, \bibinfo {author} {\bibfnamefont {H.}~\bibnamefont {Weng}}, \bibinfo
  {author} {\bibfnamefont {Z.}~\bibnamefont {Wang}}, \bibinfo {author}
  {\bibfnamefont {X.}~\bibnamefont {Dai}}, \ and\ \bibinfo {author}
  {\bibfnamefont {Z.}~\bibnamefont {Fang}},\ }\href {\doibase
  10.1103/PhysRevLett.107.186806} {\bibfield  {journal} {\bibinfo  {journal}
  {Phys. Rev. Lett.}\ }\textbf {\bibinfo {volume} {107}},\ \bibinfo {pages}
  {186806} (\bibinfo {year} {2011})}\BibitemShut {NoStop}%
\bibitem [{\citenamefont {Volovik}(2013)}]{Volovik2013}%
  \BibitemOpen
  \bibfield  {author} {\bibinfo {author} {\bibfnamefont {G.~E.}\ \bibnamefont
  {Volovik}},\ }\href {\doibase 10.1007/s10948-013-2221-5} {\bibfield
  {journal} {\bibinfo  {journal} {Journal of Superconductivity and Novel
  Magnetism}\ }\textbf {\bibinfo {volume} {26}},\ \bibinfo {pages} {2887}
  (\bibinfo {year} {2013})}\BibitemShut {NoStop}%
\bibitem [{\citenamefont {Kim}\ \emph {et~al.}(2015)\citenamefont {Kim},
  \citenamefont {Wieder}, \citenamefont {Kane},\ and\ \citenamefont
  {Rappe}}]{Kim2015}%
  \BibitemOpen
  \bibfield  {author} {\bibinfo {author} {\bibfnamefont {Y.}~\bibnamefont
  {Kim}}, \bibinfo {author} {\bibfnamefont {B.~J.}\ \bibnamefont {Wieder}},
  \bibinfo {author} {\bibfnamefont {C.~L.}\ \bibnamefont {Kane}}, \ and\
  \bibinfo {author} {\bibfnamefont {A.~M.}\ \bibnamefont {Rappe}},\ }\href
  {\doibase 10.1103/PhysRevLett.115.036806} {\bibfield  {journal} {\bibinfo
  {journal} {Phys. Rev. Lett.}\ }\textbf {\bibinfo {volume} {115}},\ \bibinfo
  {pages} {036806} (\bibinfo {year} {2015})}\BibitemShut {NoStop}%
\bibitem [{\citenamefont {Herak}\ \emph {et~al.}(2005)\citenamefont {Herak},
  \citenamefont {Berger}, \citenamefont {Prester}, \citenamefont {Miljak},
  \citenamefont {Živković}, \citenamefont {Milat}, \citenamefont {Drobac},
  \citenamefont {Popović},\ and\ \citenamefont {Zaharko}}]{Herak2005}%
  \BibitemOpen
  \bibfield  {author} {\bibinfo {author} {\bibfnamefont {M.}~\bibnamefont
  {Herak}}, \bibinfo {author} {\bibfnamefont {H.}~\bibnamefont {Berger}},
  \bibinfo {author} {\bibfnamefont {M.}~\bibnamefont {Prester}}, \bibinfo
  {author} {\bibfnamefont {M.}~\bibnamefont {Miljak}}, \bibinfo {author}
  {\bibfnamefont {I.}~\bibnamefont {Živković}}, \bibinfo {author}
  {\bibfnamefont {O.}~\bibnamefont {Milat}}, \bibinfo {author} {\bibfnamefont
  {D.}~\bibnamefont {Drobac}}, \bibinfo {author} {\bibfnamefont
  {S.}~\bibnamefont {Popović}}, \ and\ \bibinfo {author} {\bibfnamefont
  {O.}~\bibnamefont {Zaharko}},\ }\href
  {http://stacks.iop.org/0953-8984/17/i=48/a=017} {\bibfield  {journal}
  {\bibinfo  {journal} {Journal of Physics: Condensed Matter}\ }\textbf
  {\bibinfo {volume} {17}},\ \bibinfo {pages} {7667} (\bibinfo {year}
  {2005})}\BibitemShut {NoStop}%
\bibitem [{\citenamefont {Choi}\ \emph {et~al.}(2008)\citenamefont {Choi},
  \citenamefont {Lemmens}, \citenamefont {Choi},\ and\ \citenamefont
  {Berger}}]{Choi2008}%
  \BibitemOpen
  \bibfield  {author} {\bibinfo {author} {\bibfnamefont {K.~Y.}\ \bibnamefont
  {Choi}}, \bibinfo {author} {\bibfnamefont {P.}~\bibnamefont {Lemmens}},
  \bibinfo {author} {\bibfnamefont {E.~S.}\ \bibnamefont {Choi}}, \ and\
  \bibinfo {author} {\bibfnamefont {H.}~\bibnamefont {Berger}},\ }\href
  {http://stacks.iop.org/0953-8984/20/i=50/a=505214} {\bibfield  {journal}
  {\bibinfo  {journal} {Journal of Physics: Condensed Matter}\ }\textbf
  {\bibinfo {volume} {20}},\ \bibinfo {pages} {505214} (\bibinfo {year}
  {2008})}\BibitemShut {NoStop}%
\bibitem [{\citenamefont {Li}\ \emph {et~al.}()\citenamefont {Li},
  \citenamefont {Li}, \citenamefont {Hu}, \citenamefont {Li},\ and\
  \citenamefont {Fang}}]{SupMat}%
  \BibitemOpen
  \bibfield  {author} {\bibinfo {author} {\bibfnamefont {K.}~\bibnamefont
  {Li}}, \bibinfo {author} {\bibfnamefont {C.}~\bibnamefont {Li}}, \bibinfo
  {author} {\bibfnamefont {J.}~\bibnamefont {Hu}}, \bibinfo {author}
  {\bibfnamefont {Y.}~\bibnamefont {Li}}, \ and\ \bibinfo {author}
  {\bibfnamefont {C.}~\bibnamefont {Fang}},\ }\href@noop {} {\bibinfo
  {journal} {Supplementary Material, which includes Refs.\cite{Pearson, Xu2015, Lopez1984, Lopez1985}}\ }\BibitemShut {NoStop}%
\bibitem [{\citenamefont {Mansson}\ \emph {et~al.}(2012)\citenamefont
  {Mansson}, \citenamefont {Prsa}, \citenamefont {Sugiyama}, \citenamefont
  {Andreica}, \citenamefont {Luetkens},\ and\ \citenamefont
  {Berger}}]{Mansson2012}%
  \BibitemOpen
\bibfield  {journal} {  }\bibfield  {author} {\bibinfo {author} {\bibfnamefont
  {M.}~\bibnamefont {Mansson}}, \bibinfo {author} {\bibfnamefont
  {K.}~\bibnamefont {Prsa}}, \bibinfo {author} {\bibfnamefont {J.}~\bibnamefont
  {Sugiyama}}, \bibinfo {author} {\bibfnamefont {D.}~\bibnamefont {Andreica}},
  \bibinfo {author} {\bibfnamefont {H.}~\bibnamefont {Luetkens}}, \ and\
  \bibinfo {author} {\bibfnamefont {H.}~\bibnamefont {Berger}},\ }\href
  {\doibase http://dx.doi.org/10.1016/j.phpro.2012.04.059} {\bibfield
  {journal} {\bibinfo  {journal} {Physics Procedia}\ }\textbf {\bibinfo
  {volume} {30}},\ \bibinfo {pages} {142 } (\bibinfo {year}
  {2012})}\BibitemShut {NoStop}%
\bibitem [{\citenamefont {Maslen}\ \emph {et~al.}(1996)\citenamefont {Maslen},
  \citenamefont {Streltsov},\ and\ \citenamefont {Ishizawa}}]{Maslen1996}%
  \BibitemOpen
  \bibfield  {author} {\bibinfo {author} {\bibfnamefont {E.~N.}\ \bibnamefont
  {Maslen}}, \bibinfo {author} {\bibfnamefont {V.~A.}\ \bibnamefont
  {Streltsov}}, \ and\ \bibinfo {author} {\bibfnamefont {N.}~\bibnamefont
  {Ishizawa}},\ }\href {\doibase 10.1107/S0108768195013371} {\bibfield
  {journal} {\bibinfo  {journal} {Acta Crystallographica Section B}\ }\textbf
  {\bibinfo {volume} {52}},\ \bibinfo {pages} {414} (\bibinfo {year}
  {1996})}\BibitemShut {NoStop}%
\bibitem [{\citenamefont {Antic}\ \emph {et~al.}(1995)\citenamefont {Antic},
  \citenamefont {Mitric.},\ and\ \citenamefont {Rodic}}]{Antic1995}%
  \BibitemOpen
  \bibfield  {author} {\bibinfo {author} {\bibfnamefont {B.}~\bibnamefont
  {Antic}}, \bibinfo {author} {\bibfnamefont {M.}~\bibnamefont {Mitric.}}, \
  and\ \bibinfo {author} {\bibfnamefont {D.}~\bibnamefont {Rodic}},\
  }\href@noop {} {\bibfield  {journal} {\bibinfo  {journal} {Journal of
  Magnetism and Magnetic Materials}\ }\textbf {\bibinfo {volume} {145}},\
  \bibinfo {pages} {349} (\bibinfo {year} {1995})}\BibitemShut {NoStop}%
  \bibitem [{\citenamefont {Villars}\ \emph {et~al.}(2017/18) \citenamefont {Villars}, \citenamefont
  {Cenzual}}]{Pearson}%
  \BibitemOpen
  \bibfield  {author} {\bibinfo {author} {\bibfnamefont {P.}\
  \bibnamefont {Villars}}, \bibinfo {author} {\bibfnamefont {K.}\
  \bibnamefont {Cenzual}},} \href
  {http://www.crystalimpact.com/pcd/howto.htm} {\bibfield  {journal}
  {\bibinfo  {journal} {Pearson's Crystal Data: Crystal Structure Database for Inorganic Compounds (on DVD), ASM International®, Materials Park, Ohio, USA}\ }(\bibinfo {year}
  {Release 2017/18})}\BibitemShut {NoStop}%
\bibitem [{\citenamefont {Xu}\ \emph {et~al.}(2015)\citenamefont {Xu},
  \citenamefont {Liu}, \citenamefont {Kushwaha}, \citenamefont {Sankar},
  \citenamefont {Krizan}, \citenamefont {Belopolski}, \citenamefont {Neupane},
  \citenamefont {Bian}, \citenamefont {Alidoust}, \citenamefont {Chang} \emph
  {et~al.}}]{Xu2015}%
  \BibitemOpen
  \bibfield  {author} {\bibinfo {author} {\bibfnamefont {S.-Y.}\ \bibnamefont
  {Xu}}, \bibinfo {author} {\bibfnamefont {C.}~\bibnamefont {Liu}}, \bibinfo
  {author} {\bibfnamefont {S.~K.}\ \bibnamefont {Kushwaha}}, \bibinfo {author}
  {\bibfnamefont {R.}~\bibnamefont {Sankar}}, \bibinfo {author} {\bibfnamefont
  {J.~W.}\ \bibnamefont {Krizan}}, \bibinfo {author} {\bibfnamefont
  {I.}~\bibnamefont {Belopolski}}, \bibinfo {author} {\bibfnamefont
  {M.}~\bibnamefont {Neupane}}, \bibinfo {author} {\bibfnamefont
  {G.}~\bibnamefont {Bian}}, \bibinfo {author} {\bibfnamefont {N.}~\bibnamefont
  {Alidoust}}, \bibinfo {author} {\bibfnamefont {T.-R.}\ \bibnamefont {Chang}},
   \emph {et~al.},\ }\href@noop {} {\bibfield  {journal} {\bibinfo  {journal}
  {Science}\ }\textbf {\bibinfo {volume} {347}},\ \bibinfo {pages} {294}
  (\bibinfo {year} {2015})}\BibitemShut {NoStop}%
\bibitem [{\citenamefont {Sancho}\ \emph {et~al.}(1984)\citenamefont {Sancho},
  \citenamefont {Sancho},\ and\ \citenamefont {Rubio}}]{Lopez1984}%
  \BibitemOpen
  \bibfield  {author} {\bibinfo {author} {\bibfnamefont {M.~P.~L.}\
  \bibnamefont {Sancho}}, \bibinfo {author} {\bibfnamefont {J.~M.~L.}\
  \bibnamefont {Sancho}}, \ and\ \bibinfo {author} {\bibfnamefont
  {J.}~\bibnamefont {Rubio}},\ }\href
  {http://stacks.iop.org/0305-4608/14/i=5/a=016} {\bibfield  {journal}
  {\bibinfo  {journal} {Journal of Physics F: Metal Physics}\ }\textbf
  {\bibinfo {volume} {14}},\ \bibinfo {pages} {1205} (\bibinfo {year}
  {1984})}\BibitemShut {NoStop}%
\bibitem [{\citenamefont {Sancho}\ \emph {et~al.}(1985)\citenamefont {Sancho},
  \citenamefont {Sancho}, \citenamefont {Sancho},\ and\ \citenamefont
  {Rubio}}]{Lopez1985}%
  \BibitemOpen
  \bibfield  {author} {\bibinfo {author} {\bibfnamefont {M.~P.~L.}\
  \bibnamefont {Sancho}}, \bibinfo {author} {\bibfnamefont {J.~M.~L.}\
  \bibnamefont {Sancho}}, \bibinfo {author} {\bibfnamefont {J.~M.~L.}\
  \bibnamefont {Sancho}}, \ and\ \bibinfo {author} {\bibfnamefont
  {J.}~\bibnamefont {Rubio}},\ }\href
  {http://stacks.iop.org/0305-4608/15/i=4/a=009} {\bibfield  {journal}
  {\bibinfo  {journal} {Journal of Physics F: Metal Physics}\ }\textbf
  {\bibinfo {volume} {15}},\ \bibinfo {pages} {851} (\bibinfo {year}
  {1985})}\BibitemShut {NoStop}%
\end{thebibliography}

\begin{thebibliography}{6}%
\makeatletter
\providecommand \@ifxundefined [1]{%
 \@ifx{#1\undefined}
}%
\providecommand \@ifnum [1]{%
 \ifnum #1\expandafter \@firstoftwo
 \else \expandafter \@secondoftwo
 \fi
}%
\providecommand \@ifx [1]{%
 \ifx #1\expandafter \@firstoftwo
 \else \expandafter \@secondoftwo
 \fi
}%
\providecommand \natexlab [1]{#1}%
\providecommand \enquote  [1]{``#1''}%
\providecommand \bibnamefont  [1]{#1}%
\providecommand \bibfnamefont [1]{#1}%
\providecommand \citenamefont [1]{#1}%
\providecommand \href@noop [0]{\@secondoftwo}%
\providecommand \href [0]{\begingroup \@sanitize@url \@href}%
\providecommand \@href[1]{\@@startlink{#1}\@@href}%
\providecommand \@@href[1]{\endgroup#1\@@endlink}%
\providecommand \@sanitize@url [0]{\catcode `\\12\catcode `\$12\catcode
  `\&12\catcode `\#12\catcode `\^12\catcode `\_12\catcode `\%12\relax}%
\providecommand \@@startlink[1]{}%
\providecommand \@@endlink[0]{}%
\providecommand \url  [0]{\begingroup\@sanitize@url \@url }%
\providecommand \@url [1]{\endgroup\@href {#1}{\urlprefix }}%
\providecommand \urlprefix  [0]{URL }%
\providecommand \Eprint [0]{\href }%
\providecommand \doibase [0]{http://dx.doi.org/}%
\providecommand \selectlanguage [0]{\@gobble}%
\providecommand \bibinfo  [0]{\@secondoftwo}%
\providecommand \bibfield  [0]{\@secondoftwo}%
\providecommand \translation [1]{[#1]}%
\providecommand \BibitemOpen [0]{}%
\providecommand \bibitemStop [0]{}%
\providecommand \bibitemNoStop [0]{.\EOS\space}%
\providecommand \EOS [0]{\spacefactor3000\relax}%
\providecommand \BibitemShut  [1]{\csname bibitem#1\endcsname}%
\let\auto@bib@innerbib\@empty
   \bibitem [{\citenamefont {Villars}\ \emph {et~al.}(2017/18) \citenamefont {Villars}, \citenamefont
  {Cenzual}}]{Pearson2}%
  \BibitemOpen
  \bibfield  {author} {\bibinfo {author} {\bibfnamefont {P.}\
  \bibnamefont {Villars}}, \bibinfo {author} {\bibfnamefont {K.}\
  \bibnamefont {Cenzual}},} \href
  {http://www.crystalimpact.com/pcd/howto.htm} {\bibfield  {journal}
  {\bibinfo  {journal} {Pearson's Crystal Data: Crystal Structure Database for Inorganic Compounds (on DVD), ASM International®, Materials Park, Ohio, USA}\ }(\bibinfo {year}
  {Release 2017/18})}\BibitemShut {NoStop}%
\bibitem [{\citenamefont {Fang}\ \emph {et~al.}(2015)\citenamefont {Fang},
  \citenamefont {Chen}, \citenamefont {Kee},\ and\ \citenamefont
  {Fu}}]{Fang2015a2}%
  \BibitemOpen
  \bibfield  {author} {\bibinfo {author} {\bibfnamefont {C.}~\bibnamefont
  {Fang}}, \bibinfo {author} {\bibfnamefont {Y.}~\bibnamefont {Chen}}, \bibinfo
  {author} {\bibfnamefont {H.-Y.}\ \bibnamefont {Kee}}, \ and\ \bibinfo
  {author} {\bibfnamefont {L.}~\bibnamefont {Fu}},\ }\href {\doibase
  10.1103/PhysRevB.92.081201} {\bibfield  {journal} {\bibinfo  {journal} {Phys.
  Rev. B}\ }\textbf {\bibinfo {volume} {92}},\ \bibinfo {pages} {081201}
  (\bibinfo {year} {2015})}\BibitemShut {NoStop}%
\bibitem [{\citenamefont {Fang}\ \emph {et~al.}(2016)\citenamefont {Fang},
  \citenamefont {Lu}, \citenamefont {Liu},\ and\ \citenamefont
  {Fu}}]{Fang20162}%
  \BibitemOpen
  \bibfield  {author} {\bibinfo {author} {\bibfnamefont {C.}~\bibnamefont
  {Fang}}, \bibinfo {author} {\bibfnamefont {L.}~\bibnamefont {Lu}}, \bibinfo
  {author} {\bibfnamefont {J.}~\bibnamefont {Liu}}, \ and\ \bibinfo {author}
  {\bibfnamefont {L.}~\bibnamefont {Fu}},\ }\href
  {http://dx.doi.org/10.1038/nphys3782} {\bibfield  {journal} {\bibinfo
  {journal} {Nat Phys}\ }\textbf {\bibinfo {volume} {12}},\ \bibinfo {pages}
  {936} (\bibinfo {year} {2016})}\BibitemShut {NoStop}%
\bibitem [{\citenamefont {Wang}\ \emph {et~al.}(2012)\citenamefont {Wang},
  \citenamefont {Sun}, \citenamefont {Chen}, \citenamefont {Franchini},
  \citenamefont {Xu}, \citenamefont {Weng}, \citenamefont {Dai},\ and\
  \citenamefont {Fang}}]{Wang20122}%
  \BibitemOpen
  \bibfield  {author} {\bibinfo {author} {\bibfnamefont {Z.}~\bibnamefont
  {Wang}}, \bibinfo {author} {\bibfnamefont {Y.}~\bibnamefont {Sun}}, \bibinfo
  {author} {\bibfnamefont {X.-Q.}\ \bibnamefont {Chen}}, \bibinfo {author}
  {\bibfnamefont {C.}~\bibnamefont {Franchini}}, \bibinfo {author}
  {\bibfnamefont {G.}~\bibnamefont {Xu}}, \bibinfo {author} {\bibfnamefont
  {H.}~\bibnamefont {Weng}}, \bibinfo {author} {\bibfnamefont {X.}~\bibnamefont
  {Dai}}, \ and\ \bibinfo {author} {\bibfnamefont {Z.}~\bibnamefont {Fang}},\
  }\href {\doibase {10.1103/PhysRevB.85.195320}} {\bibfield  {journal}
  {\bibinfo  {journal} {Phys. Rev. B}\ }\textbf {\bibinfo {volume} {{85}}},\
  \bibinfo {pages} {195320} (\bibinfo {year} {{2012}})}\BibitemShut {NoStop}%
\bibitem [{\citenamefont {Xu}\ \emph {et~al.}(2015)\citenamefont {Xu},
  \citenamefont {Liu}, \citenamefont {Kushwaha}, \citenamefont {Sankar},
  \citenamefont {Krizan}, \citenamefont {Belopolski}, \citenamefont {Neupane},
  \citenamefont {Bian}, \citenamefont {Alidoust}, \citenamefont {Chang} \emph
  {et~al.}}]{Xu20152}%
  \BibitemOpen
  \bibfield  {author} {\bibinfo {author} {\bibfnamefont {S.-Y.}\ \bibnamefont
  {Xu}}, \bibinfo {author} {\bibfnamefont {C.}~\bibnamefont {Liu}}, \bibinfo
  {author} {\bibfnamefont {S.~K.}\ \bibnamefont {Kushwaha}}, \bibinfo {author}
  {\bibfnamefont {R.}~\bibnamefont {Sankar}}, \bibinfo {author} {\bibfnamefont
  {J.~W.}\ \bibnamefont {Krizan}}, \bibinfo {author} {\bibfnamefont
  {I.}~\bibnamefont {Belopolski}}, \bibinfo {author} {\bibfnamefont
  {M.}~\bibnamefont {Neupane}}, \bibinfo {author} {\bibfnamefont
  {G.}~\bibnamefont {Bian}}, \bibinfo {author} {\bibfnamefont {N.}~\bibnamefont
  {Alidoust}}, \bibinfo {author} {\bibfnamefont {T.-R.}\ \bibnamefont {Chang}},
   \emph {et~al.},\ }\href@noop {} {\bibfield  {journal} {\bibinfo  {journal}
  {Science}\ }\textbf {\bibinfo {volume} {347}},\ \bibinfo {pages} {294}
  (\bibinfo {year} {2015})}\BibitemShut {NoStop}%
\bibitem [{\citenamefont {Sancho}\ \emph {et~al.}(1984)\citenamefont {Sancho},
  \citenamefont {Sancho},\ and\ \citenamefont {Rubio}}]{Lopez19842}%
  \BibitemOpen
  \bibfield  {author} {\bibinfo {author} {\bibfnamefont {M.~P.~L.}\
  \bibnamefont {Sancho}}, \bibinfo {author} {\bibfnamefont {J.~M.~L.}\
  \bibnamefont {Sancho}}, \ and\ \bibinfo {author} {\bibfnamefont
  {J.}~\bibnamefont {Rubio}},\ }\href
  {http://stacks.iop.org/0305-4608/14/i=5/a=016} {\bibfield  {journal}
  {\bibinfo  {journal} {Journal of Physics F: Metal Physics}\ }\textbf
  {\bibinfo {volume} {14}},\ \bibinfo {pages} {1205} (\bibinfo {year}
  {1984})}\BibitemShut {NoStop}%
\bibitem [{\citenamefont {Sancho}\ \emph {et~al.}(1985)\citenamefont {Sancho},
  \citenamefont {Sancho}, \citenamefont {Sancho},\ and\ \citenamefont
  {Rubio}}]{Lopez19852}%
  \BibitemOpen
  \bibfield  {author} {\bibinfo {author} {\bibfnamefont {M.~P.~L.}\
  \bibnamefont {Sancho}}, \bibinfo {author} {\bibfnamefont {J.~M.~L.}\
  \bibnamefont {Sancho}}, \bibinfo {author} {\bibfnamefont {J.~M.~L.}\
  \bibnamefont {Sancho}}, \ and\ \bibinfo {author} {\bibfnamefont
  {J.}~\bibnamefont {Rubio}},\ }\href
  {http://stacks.iop.org/0305-4608/15/i=4/a=009} {\bibfield  {journal}
  {\bibinfo  {journal} {Journal of Physics F: Metal Physics}\ }\textbf
  {\bibinfo {volume} {15}},\ \bibinfo {pages} {851} (\bibinfo {year}
  {1985})}\BibitemShut {NoStop}%

\end{thebibliography}
\end{document}